\documentclass[runningheads]{llncs}

\usepackage{eccv}

\usepackage{eccvabbrv}
\usepackage{algorithmic}
\usepackage{algorithm}
\usepackage{graphicx}
\usepackage{booktabs}
\usepackage{multirow}
\usepackage{adjustbox}
\usepackage{bm}

\usepackage[accsupp]{axessibility}  

\usepackage{hyperref}

\usepackage{orcidlink}

\makeatletter
\newcommand{\printfnsymbol}[1]{%
  \textsuperscript{\@fnsymbol{#1}}%
}
\makeatother

\usepackage[normalem]{ulem} 

\begin{document}

\title{Skeleton Recall Loss for Connectivity Conserving and Resource Efficient Segmentation of Thin Tubular Structures} 
\titlerunning{Skeleton Recall Loss}

\author{Yannick Kirchhoff\inst{1,2,3}\orcidlink{0000-0001-8124-8435}\thanks{Contributed equally. Each co-first author may list themselves as lead author on their CV.} \and
Maximilian R. Rokuss\inst{1,2}\orcidlink{0009-0004-4560-0760}\printfnsymbol{1} \and
Saikat Roy\inst{1,2}\orcidlink{0000-0002-0809-6524}\printfnsymbol{1} \and
Balint Kovacs\inst{1,4}\orcidlink{0000-0002-1191-0646} \and
Constantin Ulrich\inst{1,4}\orcidlink{0000-0003-3002-8170} \and
Tassilo Wald\inst{1,2,5}\orcidlink{0009-0007-5222-2683} \and
Maximilian Zenk\inst{1,4}\orcidlink{0000-0002-8933-5995} \and
Philipp Vollmuth\inst{1,6,7}\orcidlink{0000-0002-6224-0064} \and
Jens Kleesiek\inst{8,9}\orcidlink{0000-0001-8686-0682} \and
Fabian Isensee\inst{1,5}\orcidlink{0000-0002-3519-5886} \and
Klaus Maier-Hein\inst{1,10}\orcidlink{0000-0002-6626-2463}
}

\authorrunning{Y. Kirchhoff, M. Rokuss, S. Roy et al.}

\institute{German Cancer Research Center (DKFZ) Heidelberg, Division of Medical Image Computing, Germany
\and
Faculty of Mathematics and Computer Science, Heidelberg University, Germany
\and
HIDSS4Health - Helmholtz Information and Data Science School for Health, Karlsruhe/Heidelberg, Germany
\and
Medical Faculty Heidelberg, Heidelberg University, Heidelberg, Germany
\and
Helmholtz Imaging, German Cancer Research Center, Heidelberg, Germany
\and
Division for Computational Radiology Clinical AI (CCIBonn.ai), Clinic for Neuroradiology, University Hospital Bonn, Bonn, Germany
\and
Medical Faculty Bonn, University of Bonn, Bonn, Germany
\and
Institute for Artificial Intelligence in Medicine (IKIM), University Hospital Essen, Essen, Germany
\and
Cancer Research Center Cologne Essen (CCCE), West German Cancer Center Essen, University Hospital Essen, Essen, Germany
\and
Pattern Analysis and Learning Group, Department of Radiation Oncology, Heidelberg University Hospital
}

\maketitle

\begin{abstract}
    Accurately segmenting thin tubular structures, such as vessels, nerves, roads or concrete cracks, is a crucial task in computer vision. Standard deep learning-based segmentation loss functions, such as Dice or Cross-Entropy, focus on volumetric overlap, often at the expense of preserving structural connectivity or topology. This can lead to segmentation errors that adversely affect downstream tasks, including flow calculation, navigation, and structural inspection. Although current topology-focused losses mark an improvement, they introduce significant computational and memory overheads. This is particularly relevant for 3D data, rendering these losses infeasible for larger volumes as well as increasingly important multi-class segmentation problems. To mitigate this, we propose a novel Skeleton Recall Loss, which effectively addresses these challenges by circumventing intensive GPU-based calculations with inexpensive CPU operations. It demonstrates overall superior performance to current state-of-the-art approaches on five public datasets for topology-preserving segmentation, while substantially reducing computational overheads by more than $90\%$. In doing so, we introduce the first multi-class capable loss function for thin structure segmentation, excelling in both efficiency and efficacy for topology-preservation. Our code is available to the community, providing a foundation for further advancements, at: \url{https://github.com/MIC-DKFZ/Skeleton-Recall}.

\end{abstract}

\section{Introduction}
\label{sec:intro}
\begin{figure}[tbp]
    \centering
        \begin{subfigure}[b]{0.19\textwidth}
        \centering
        \includegraphics[width=\textwidth, height=2.35cm]{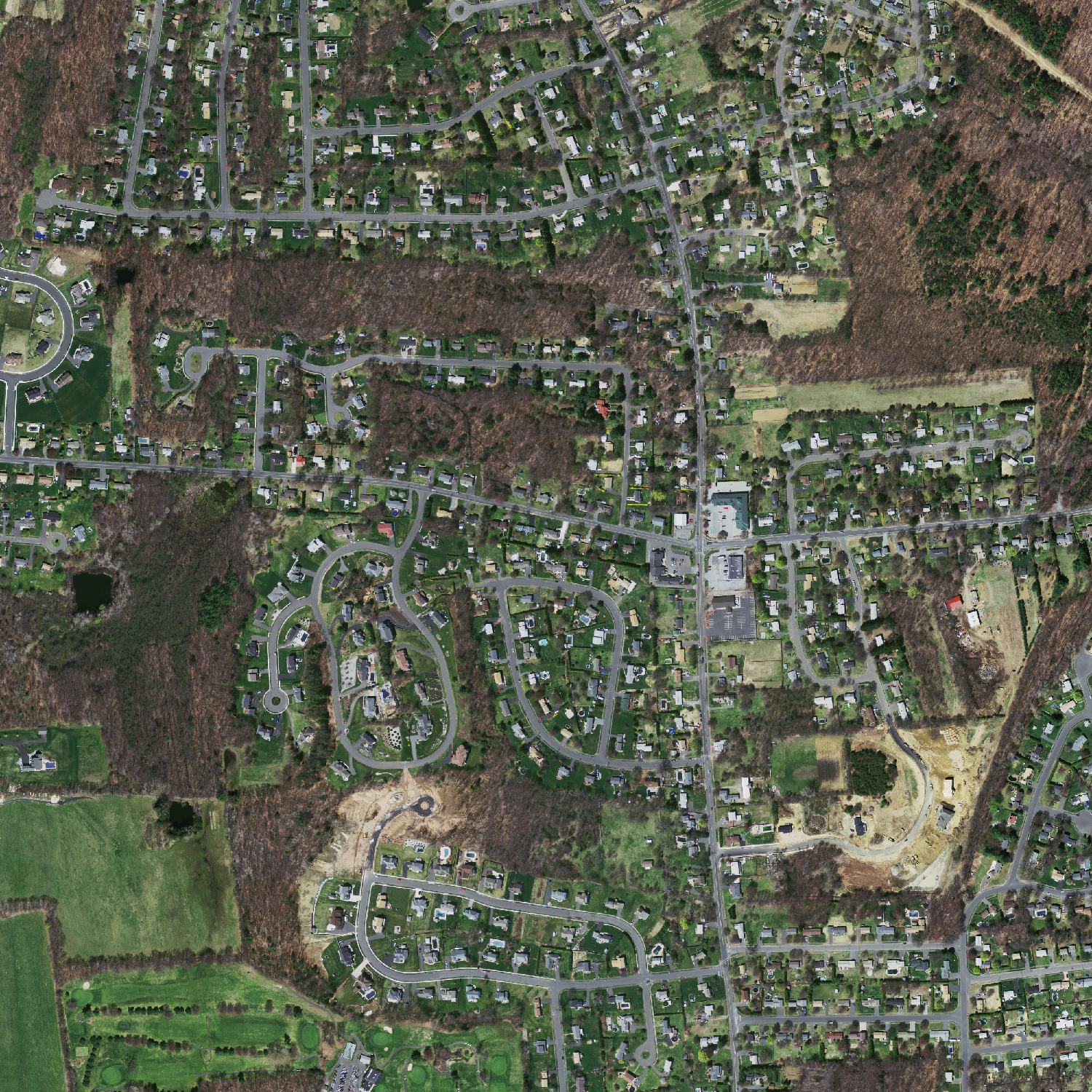}

        \includegraphics[width=\textwidth, height=2.35cm]{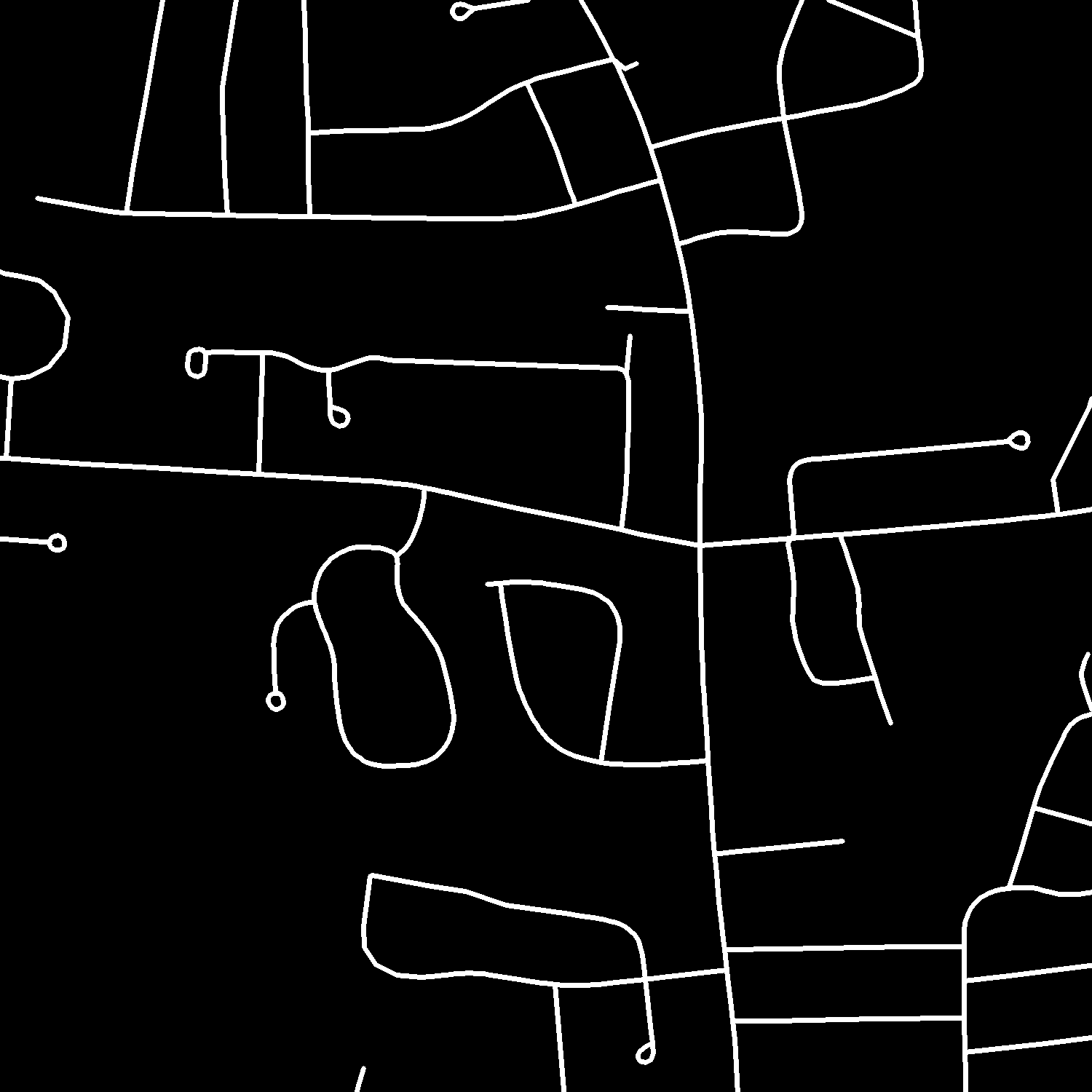}
        \caption{Roads}
        \end{subfigure}
        \begin{subfigure}[b]{0.19\textwidth}
        \centering  
        \includegraphics[width=\textwidth, height=2.35cm]{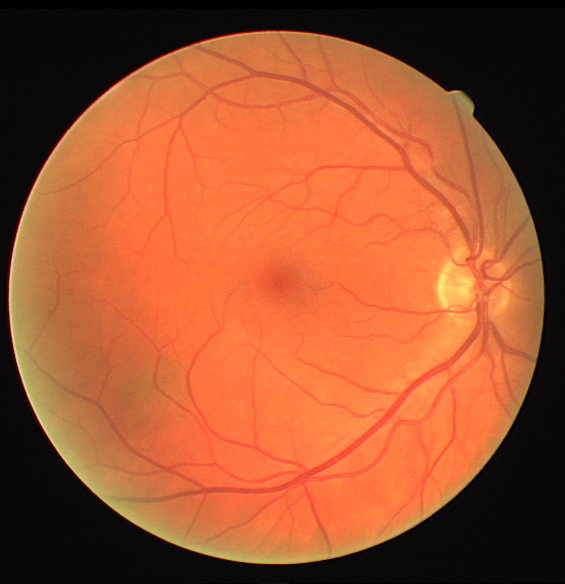}

        \includegraphics[width=\textwidth, height=2.35cm]{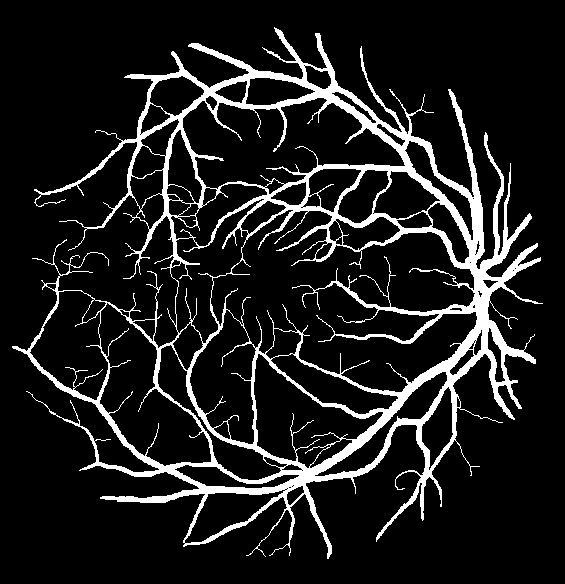}
        \caption{DRIVE}
        \end{subfigure}
        \begin{subfigure}[b]{0.19\textwidth}
        \centering
        \includegraphics[width=\textwidth, height=2.35cm]{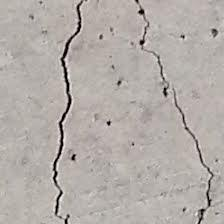}
        
        \includegraphics[width=\textwidth, height=2.35cm]{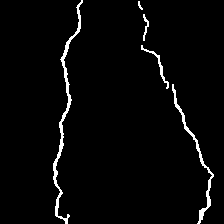}   
        \caption{Cracks}
        \end{subfigure}
        \begin{subfigure}[b]{0.19\textwidth}
        \centering
        \includegraphics[width=\textwidth, height=2.35cm]{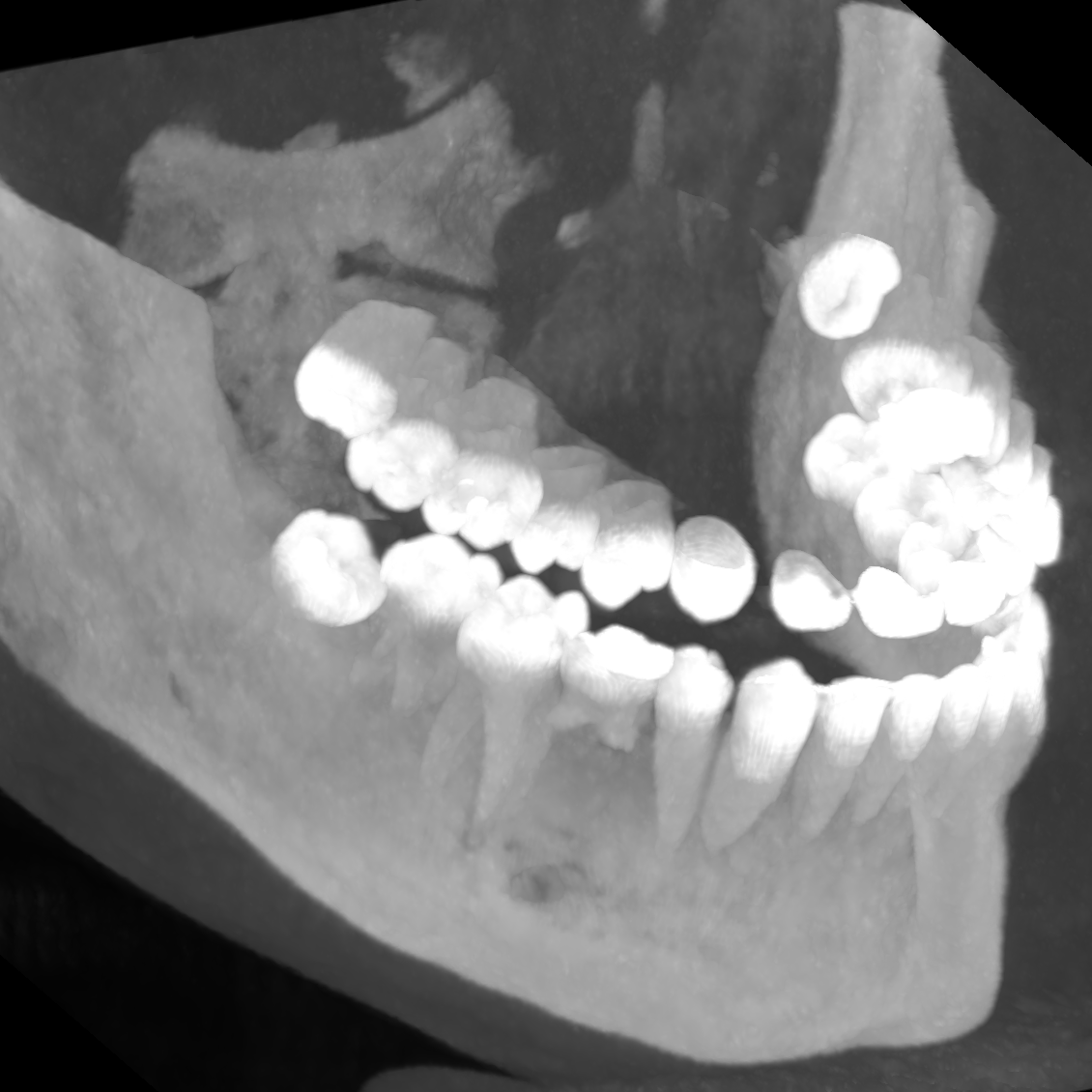}

        \includegraphics[width=\textwidth, height=2.35cm]{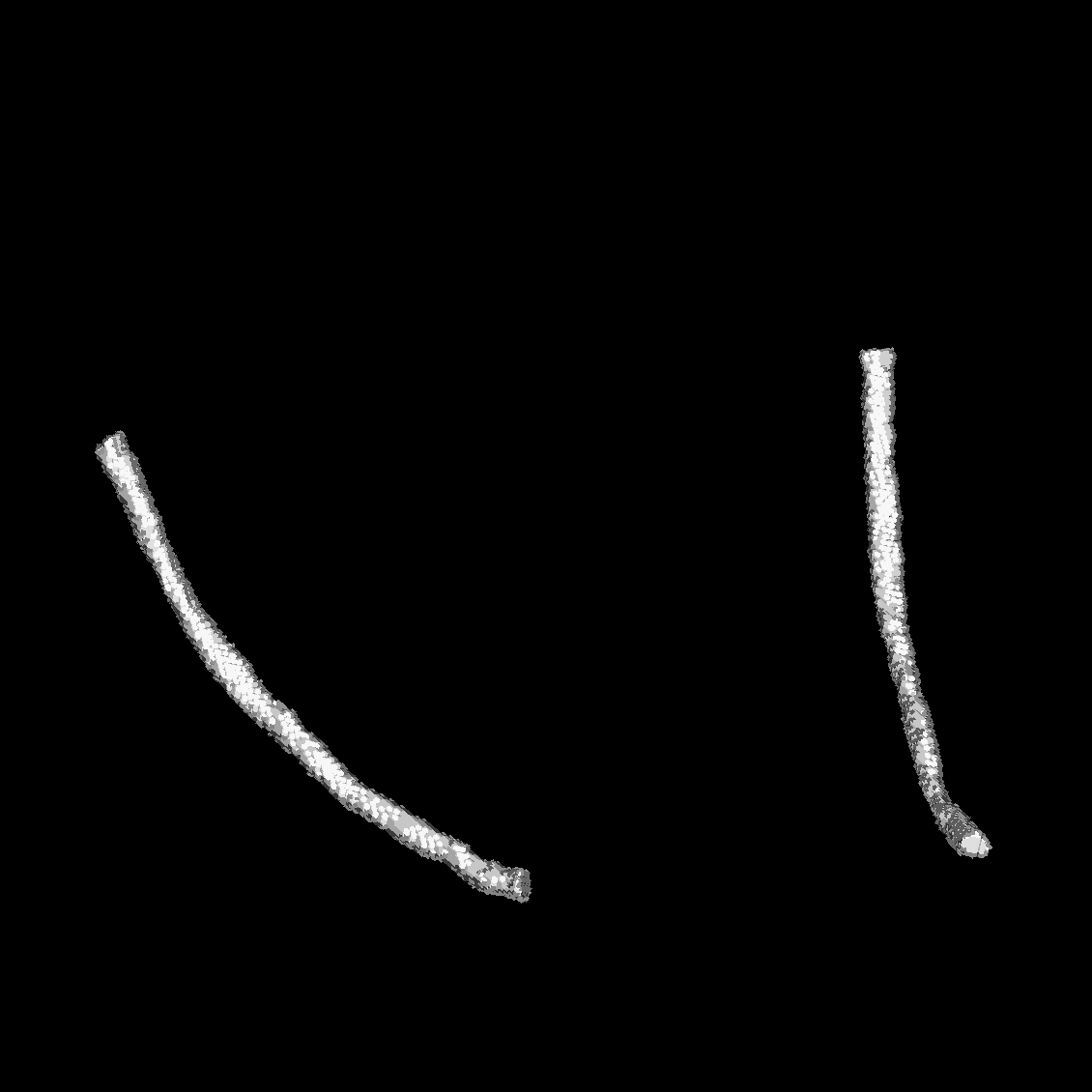}
        \caption{Toothfairy}
        \end{subfigure}
        \begin{subfigure}[b]{0.19\textwidth}
        \centering
        \includegraphics[width=\textwidth, height=2.35cm]{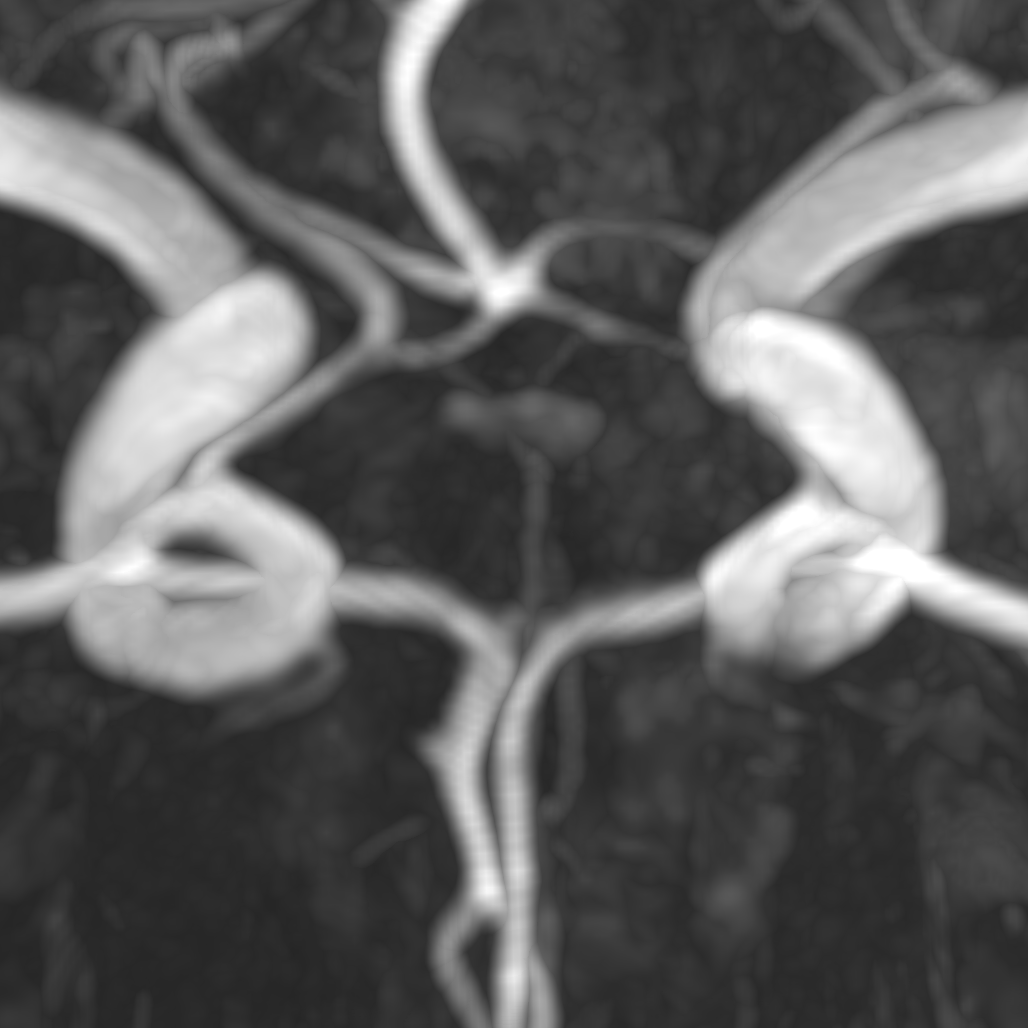}

        \includegraphics[width=\textwidth, height=2.35cm]{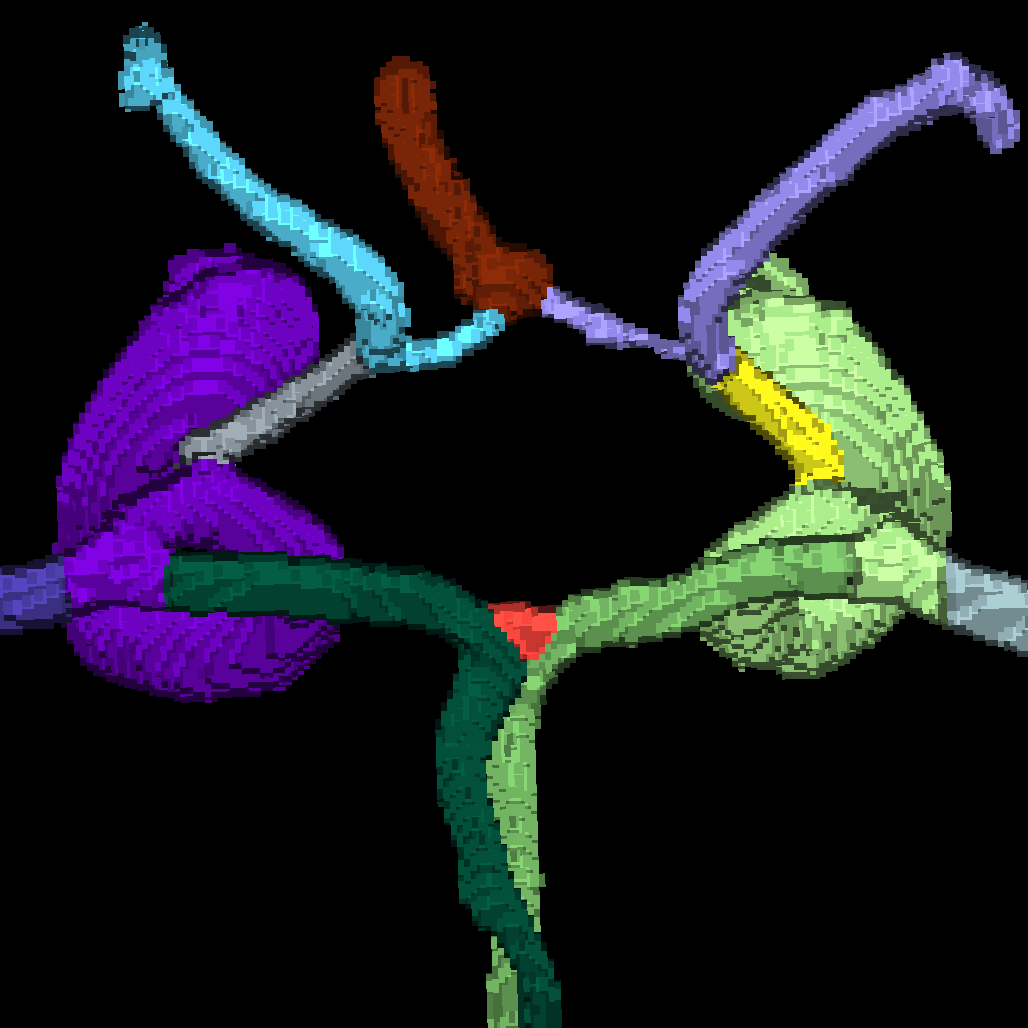}
        \caption{TopCoW}
        \end{subfigure}
    
    \caption{\textbf{Diversity of thin structures.} Segmentation of thin structures is a challenging task in engineering and medical imaging. This is highlighted in 5 diverse datasets used in this work to incorporate the segmentation of: a) Roads in satellite imagery, b) Retinal vessels, c) Cracks in concrete structures, d) Inferior alveolar canal in facial CTs, and e) Circle of Willis arterial vessel components.}
    \label{fig:prob-defn}
\end{figure}

The precise segmentation of thin tubular structures is a critical task across diverse domains in engineering and medical applications (\cref{fig:prob-defn}). Topological correctness is fundamental for facilitating downstream tasks such as analyzing blood flow dynamics, delineating neuronal boundaries in Electron Microscopy imagery, evaluating risk factors for vascular pathologies, aiding in surgical planning, and optimizing route planning~\cite{roads,arganda2015crowdsourcing, drive, topcow, toothfairycvpr, toothfairydata}. Classical approaches for the automated segmentation of thin curvilinear structures have encompassed methods including image transforms \cite{palti1997identifying, subirats2006automation}, mathematical morphologies \cite{zana2001segmentation, roychowdhury2015iterative}, filtering \cite{koller1995multiscale, hoover2000locating, lemaitre2011detection}, differential operators \cite{steger1996extracting}, among others \cite{fraz2012blood, lesage2009review, mena2003state, chambon2011automatic, bibiloni2016survey}. Deep learning based techniques have played an increasing role in recent years with standard segmentation networks such as UNet \cite{ronneberger2015u} being popular. Standard overlap-based losses (eg. dice-similarity coefficient \cite{zijdenbos1994morphometric}) enable such networks to segment large structures while often struggling with small elongated ones \cite{metricsreloaded} as shown in \cref{fig:vis_story}.

\begin{figure}[tbp]
    \centering
    \begin{subfigure}[b]{0.244\textwidth}
        \centering
        \includegraphics[width=\textwidth]{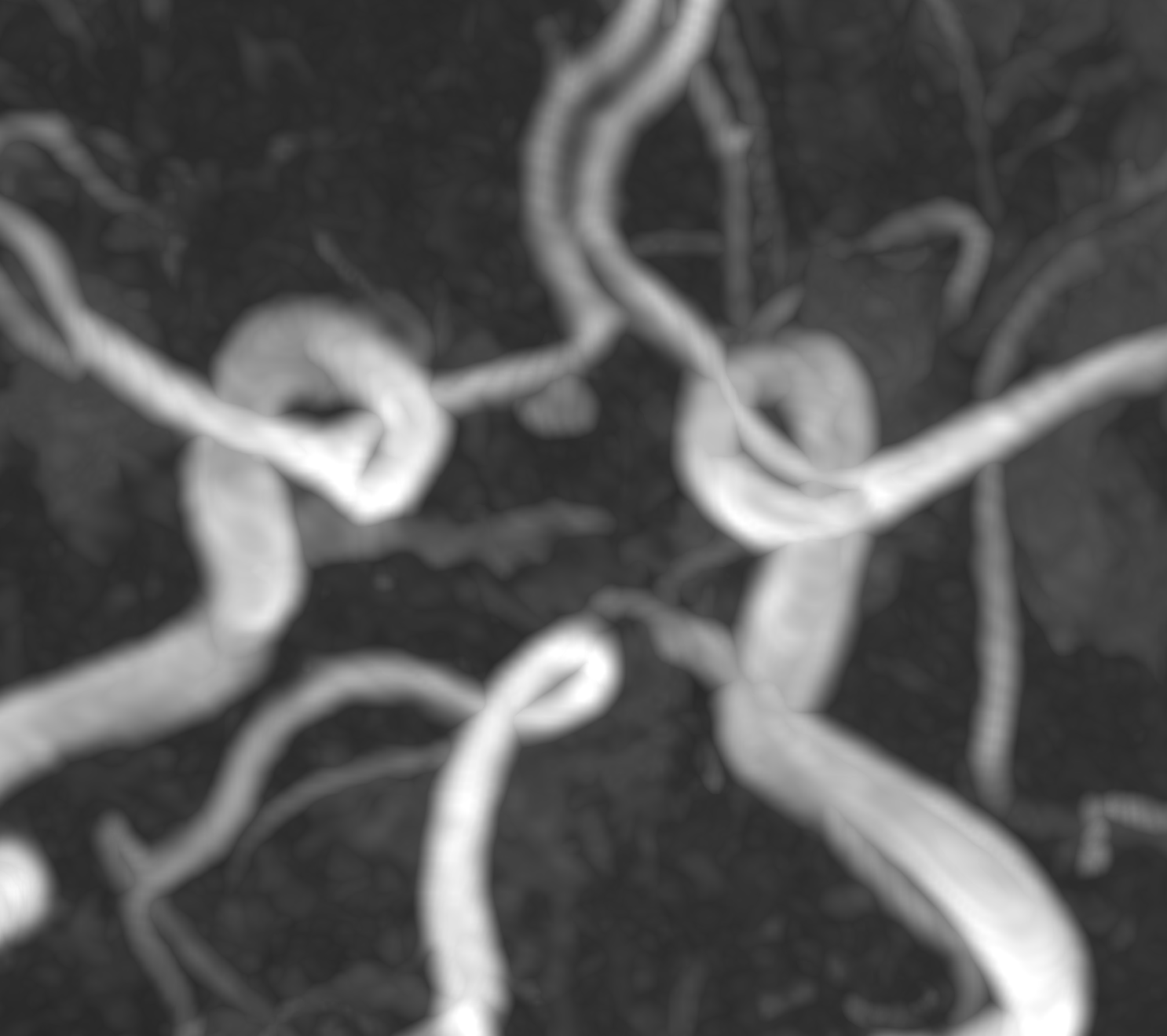}
        \caption{MRA Image}
        \label{fig:img}
    \end{subfigure}
    \hfill
    \begin{subfigure}[b]{0.244\textwidth}
        \centering
        \includegraphics[width=\textwidth]{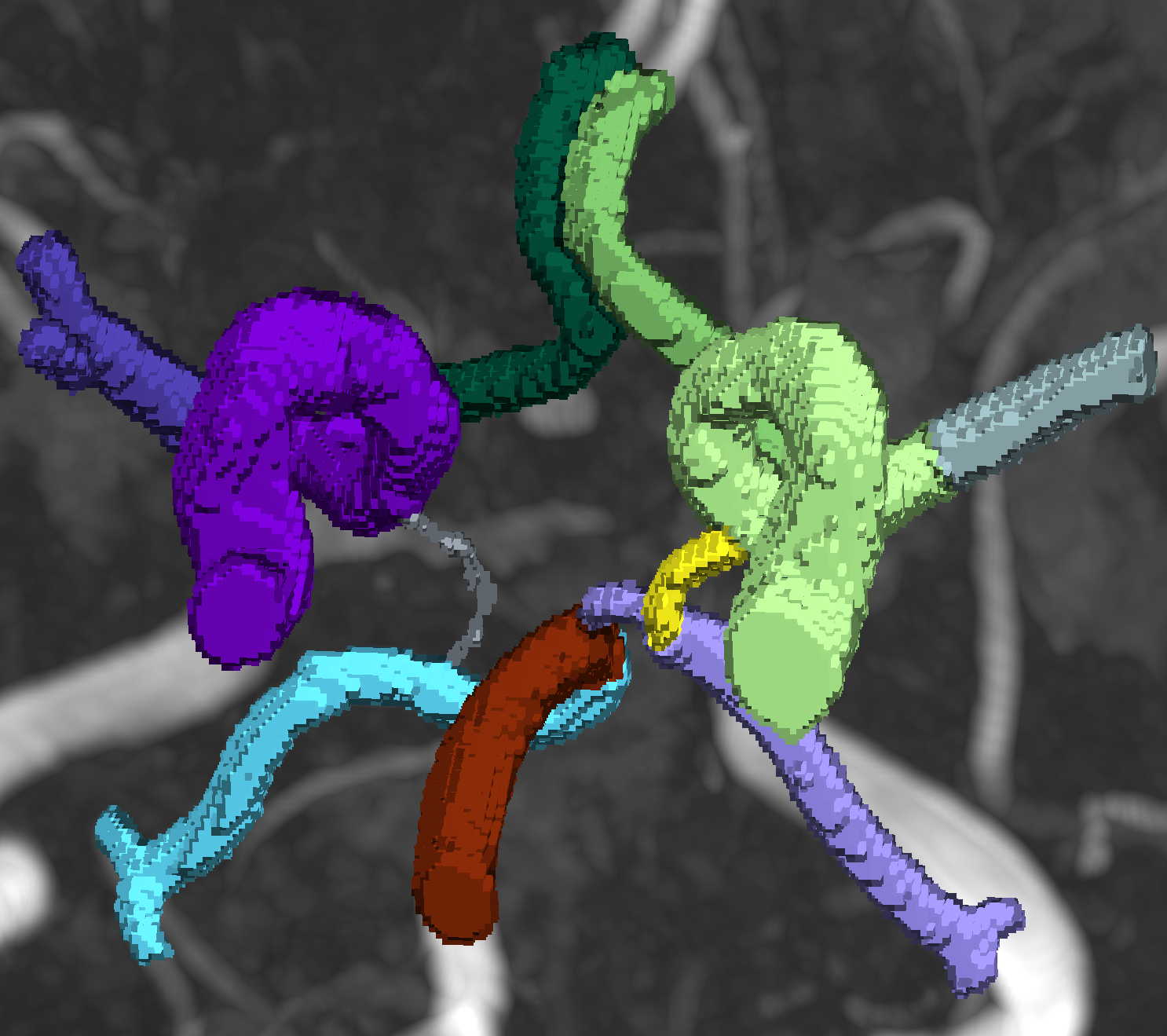}
        \caption{GT segmentation}
        \label{fig:seg}
    \end{subfigure}
    \hfill
    \begin{subfigure}[b]{0.244\textwidth}
        \centering
        \includegraphics[width=\textwidth]{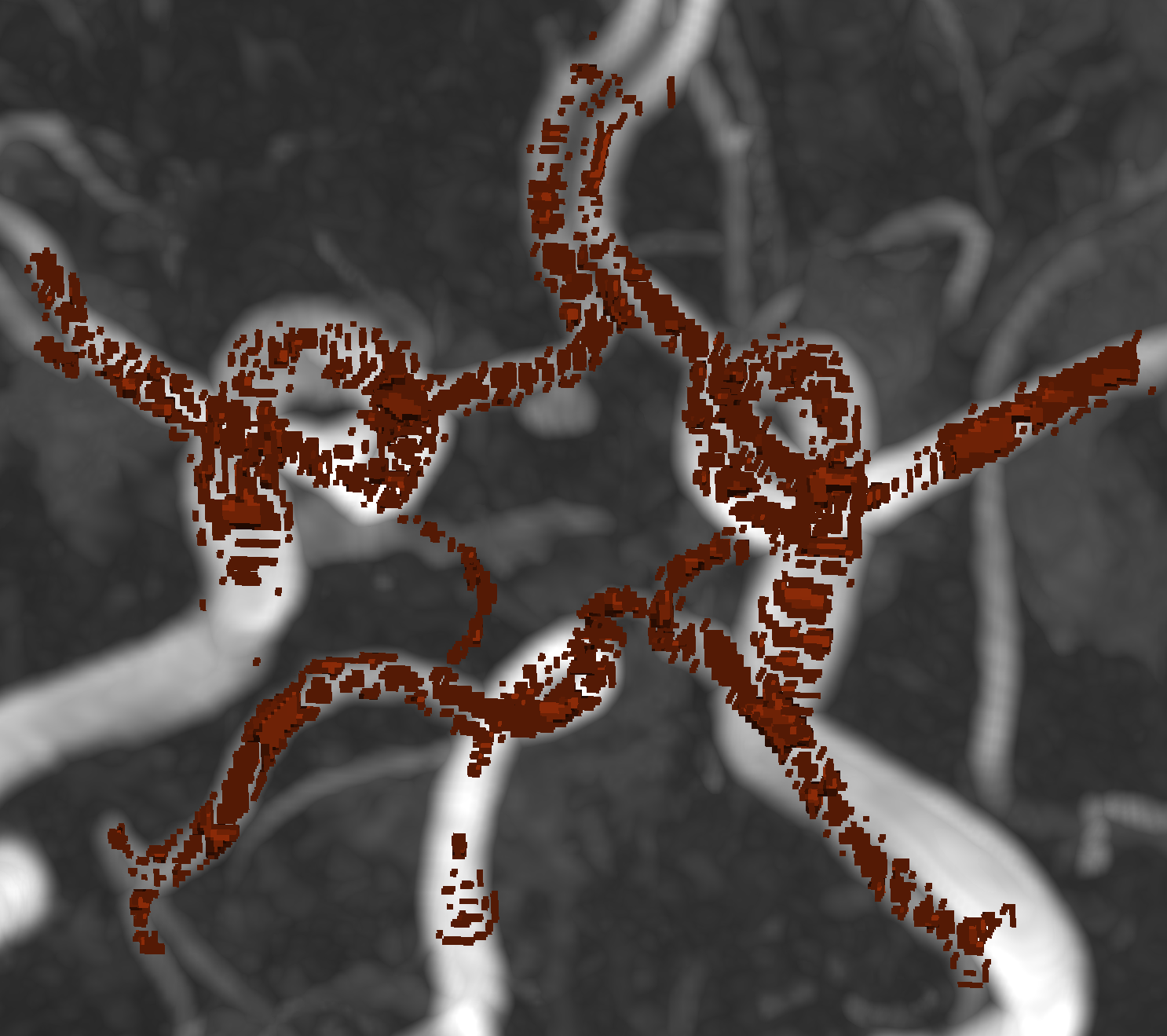}
        \caption{Soft Skeleton~\cite{clDice}}
        \label{fig:soft-skel}
    \end{subfigure}
    \hfill
    \begin{subfigure}[b]{0.244\textwidth}
        \centering
        \includegraphics[width=\textwidth]{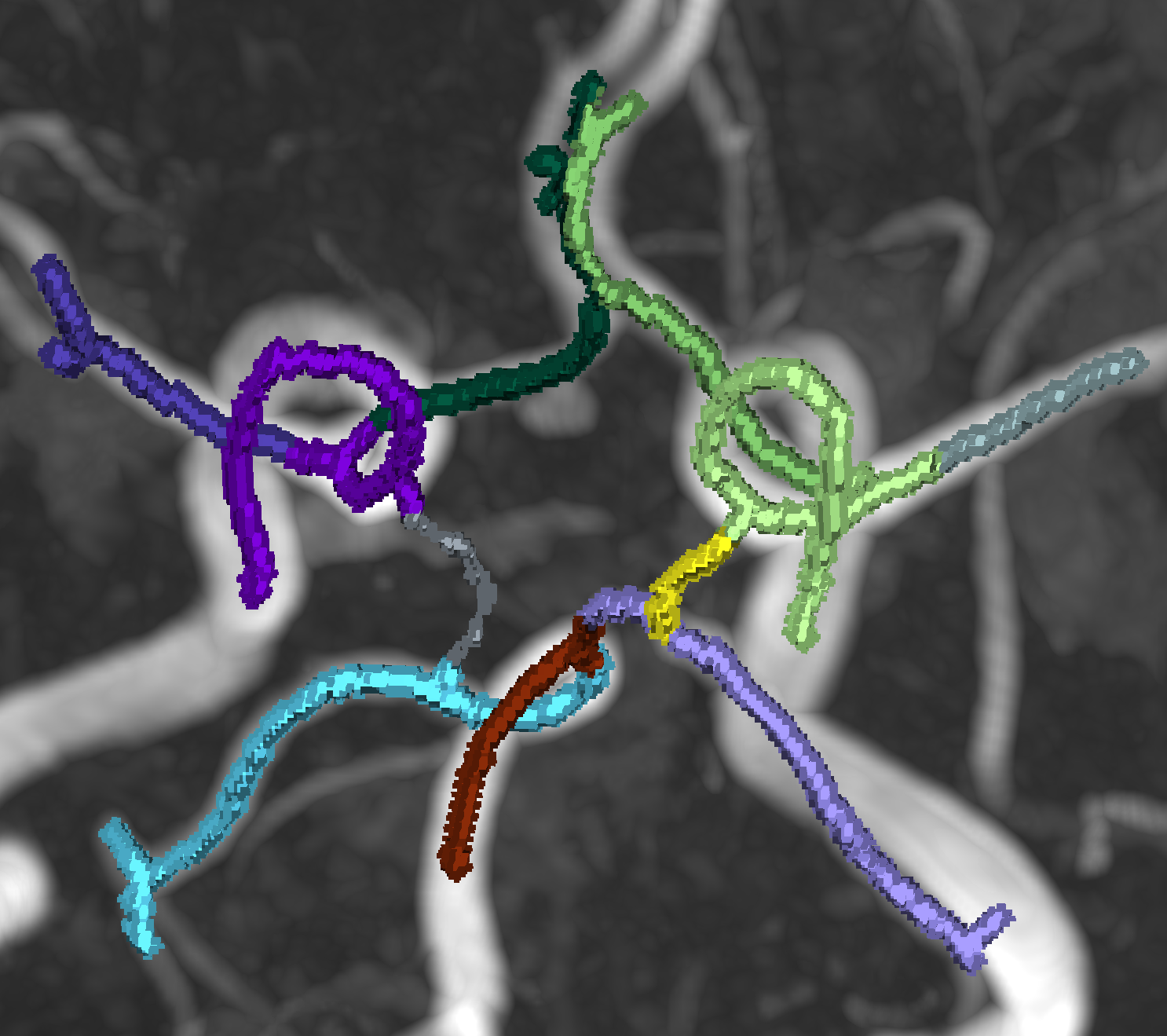}
        \caption{Tubed Skeleton}
        \label{fig:skel}
    \end{subfigure}
    \caption{\textbf{The challenges of Differentiable Skeletons.} Visual comparison of (c) the soft skeleton used for the calculation of the clDice Loss~\cite{clDice} and (d) the proposed tubed skeleton used for Skeleton Recall Loss for (a) an image and the corresponding (b) ground truth segmentation, originating from the TopCoW dataset~\cite{topcow}.}
    \label{fig:skel-comparison}
\end{figure}

Multiple methods have been introduced in recent years to address the challenges of segmenting thin curvilinear structures but are often domain-specific or require the use of specialized networks \cite{wang2019context,mou2019cs, mou2021cs2, mosinska2019joint, lin2023dtu, cheng2021joint, he2022curv}. Recently, centerlineDice~\cite{clDice} (clDice) was introduced encompassing both a \textit{loss function} and a \textit{metric} for measuring connectivity in segmentation of thin structures. Effectively, it incorporates the skeleton of a segmentation into the dice calculation. While the clDice metric uses the exact skeleton, the clDice loss works with a differentiable approximation of the skeleton. It is used to enable architecture-independent, topology-aware segmentation of thin tubular structures and is considered as state-of-the art. However, despite its advantages, it introduces a large computational overhead. Furthermore, the differentiable \textit{Soft Skeleton} used in the loss calculation can often be jagged as depicted in \cref{fig:skel-comparison}, leading to inaccuracies in segmentation. While a follow-up approach \cite{menten2023skeletonization} attempted to address this limitation by introducing a topological-correct differentiable skeleton, it did so while being even more computationally expensive. This limitation becomes particularly pronounced when dealing with large volumes or multi-class segmentation problems common to 3D medical image segmentation, rendering training on multi-class 3D datasets challenging to infeasible even on modern hardware.

In response to these challenges, we propose \textbf{\textit{Skeleton Recall Loss}}, a novel loss function tailored to address the intricacies associated with thin structures in segmentation tasks. \textit{Skeleton Recall Loss} demonstrates the following strengths:

\begin{figure}[t]
    \includegraphics[width=\textwidth]{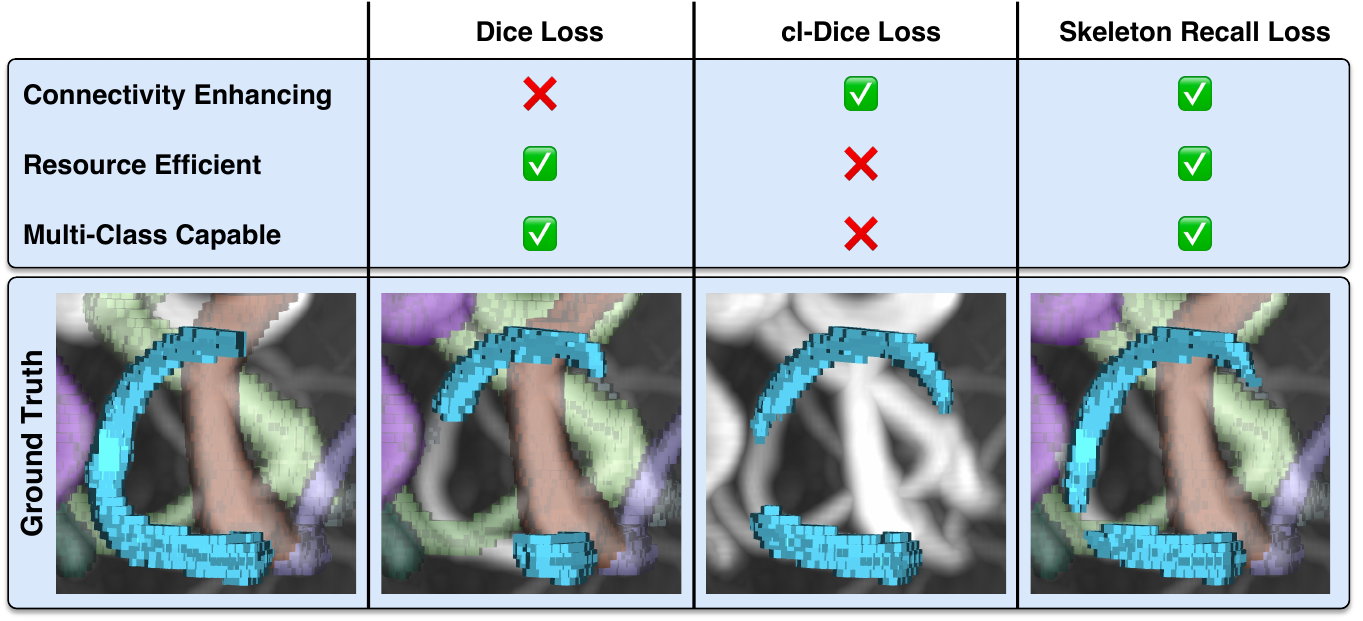}
    \caption{\textbf{Comparison of state-of-the-art loss functions on the task of thin structure segmentation.} \textit{Top:} Our Skeleton Recall Loss efficiently addresses connectivity conservation, unlike standard dice loss, without the overhead of clDice Loss, making it ideal for multi-class problems as well. \textit{Bottom:} Qualitative results on the TopCoW\cite{topcow} dataset. Due to computational cost clDice Loss can not be used for multi-class segmentation.
    }
    \label{fig:vis_story}
\end{figure}

\begin{enumerate}

    \item \textbf{Minimal training time:} The Tubed Skeleton required by \textit{Skeleton Recall Loss} can be computed with simple CPU-based operations using common image processing frameworks (\eg scikit-image \cite{van2014scikit}) as part of data-loading or even be precomputed. It is then used in a simple additional soft recall loss with the prediction, thus requiring very little additional training time.

    \item \textbf{Minimal training memory:} The utilization of \textit{Skeleton Recall Loss} entails minimal GPU Memory overhead. Unlike approaches reliant on a differentiable skeleton in prediction or ground truth during training, \textit{Skeleton Recall Loss} sidesteps the computationally taxing GPU-based skeletonization process, thus necessitating only a marginal increase in GPU training memory.

    \item \textbf{Domain and architecture agnostic:} \textit{Skeleton Recall Loss} exhibits inherent plug-and-play characteristics, seamlessly integrating into a wide array of 2D and 3D segmentation tasks without imposing architectural constraints. It operates without the need for specialized networks or modifications to underlying segmentation architectures.

    \item \textbf{Multi-class compatibility:} \textit{Skeleton Recall Loss} integrates seamlessly with multi-class labels, while competing methods like \textit{clDice Loss}, often face near insurmountable computational challenges on such problems.
\end{enumerate}

Skeleton Recall Loss yields overall superior results to a baseline network without topological losses, as well as against clDice Loss as a state-of-the-art topological loss. We demonstrate this effectiveness on extensive multi-domain evaluation on 5 publicly available datasets.
Notably, our loss function inherently feasibly supports multi-class segmentation problems and thus can be considered a new state-of-the-art for dilineating thin curvilinear structures in natural as well as medical images.

\section{Related Work}
Deep learning-based approaches for segmentation of thin curvilinear structures often involve specialized networks. In \cite{mosinska2019joint}, a joint network was trained with a shared encoder to use 2 decoders to simultaneously segment as well as score a tubular path. Similarly, in \cite{cheng2021joint}, a joint network was proposed to simultaneously learn features as well as global topology. Also, in \cite{lin2023dtu}, a pair of sequential UNets was used, the first of which performed a coarse prediction while the other detected missing or false splits in the structure. A similar end-to-end approach in \cite{mou2021cs2} used a channel and spatial attention module, which was incorporated within the bottleneck of an encoder-decoder network for segmenting thin structures in medical images. In another work \cite{peng2023curvilinear}, an oriented derivative of stick (ODoS) filter output was refined using a succession of UNets to obtain effective segmentation of curvilinear objects. 
However, alternative approaches utilize specialized loss functions to assist the network in preserving topology in segmentation outputs. In \cite{clough2020topological}, topological priors were incorporated into network training by the usage of a persistent-homology based differentiable topological prior as a loss function. Another work \cite{hu2019topology}, attempted to preserve topological information by enforcing that the prediction and ground truth have the same Betti number via a novel loss function. Recently, the use of a \textit{differentiable skeleton} has emerged as the predominant method for topology-aware segmentation of thin structures following the introduction of the centerlineDice (\textit{clDice}) loss function\cite{clDice}, outperforming persistent-homology based approaches. This method is complemented by the introduction of the clDice metric, a well-established measure of connectedness. The differentiable skeleton proposed by \cite{clDice} was improved in \cite{viti2022coronary} where a Soft-Persistent Skeleton was proposed for coronal artery tracking. Alternatively, in \cite{rouge2023cascaded}, the differentiable skeleton was predicted by a secondary network in addition to the primary segmentation output. Most recently, a topologically correct differentiable skeletonization algorithm was introduced in \cite{menten2023skeletonization}, overcoming previous approximation of skeletons, while still requiring massive computational resources to do so. 

\section{Methodology}

\subsection{The challenges of Differentiable Skeletons}

The usage of a differentiable skeleton based loss \cite{clDice,viti2022coronary,rouge2023cascaded,menten2023skeletonization} in training a deep neural network to segment thin tubular structures is an intuitive approach to preserve connectivity. However, it is fraught with challenges which can be multi-faceted in nature. One of the most easily demonstrable issues is shown in \cref{fig:skel-comparison}. As mentioned earlier, the so-called \textit{Soft skeletonization} of \textit{clDice Loss} can lead to perforated and jagged skeletons, especially in 3D, which results in inaccuracies for the clDice Loss calculation. This is in addition to the enormous GPU memory and training time overheads that are a natural part of a GPU-based differentiable skeletonization process on both the ground truth as well as the network prediction. This overhead is demonstrated in \cref{fig:mem_time_plots}, and can render effective training almost infeasible in multi-class datasets with large 3D input volumes such as TopCoW \cite{topcow} (which is used in this work) without access to significant computational resources. While follow-up work in \cite{menten2023skeletonization} allowed for a relative improvement in topologically accurate differentiable skeletonization, it further aggravated the issues with excessive resource utilization.

\subsection{Skeleton Recall Loss: Connectivity conservation on thin structures without differentiable skeletons}

\begin{figure}[t]
    \centering
    \includegraphics[width=0.9\textwidth]{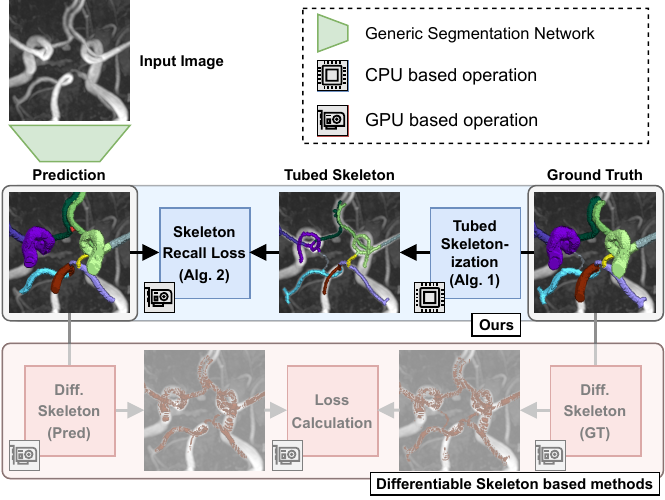}
    \caption{\textbf{Overview of our method in comparison to differentiable skeleton based approaches.} Initially, a segmentation network (\textit{green}) predicts a segmentation mask. Our proposed Skeleton Recall Loss (\textit{blue}) calculates the soft recall of the prediction on the precomputed Tubed Skeleton of the ground truth. In doing so, we mitigate the massive overheads introduced by differentiable skeleton based methods (\textit{red}).}
    \label{fig:method}
\end{figure}

Skeleton Recall Loss is a loss function designed to preserve connectivity in thin tubular structures without incurring massive computational overheads. It is universally applicable, regardless of whether the inputs are 2D or 3D. It does so by avoiding the GPU-based soft-skeletonization on the prediction and ground truth. Instead, a tubed skeletonization is performed on the ground truth, followed by a soft recall loss against the predicted segmentation output. This is illustrated in \cref{fig:method} and further discussed in the following sections.

\subsubsection{Tubed Skeletonization}
The usage of a skeleton for the preservation of connectivity is an effective method, but it does not need to be differentiable. In this work, we extract a \textit{tubed skeleton} from the ground truth as demonstrated in \cref{alg:tubed_skel}. Initially, we binarize the ground truth segmentation mask and compute its skeleton using methods outlined in \cite{zhang1984fast} for 2D and \cite{lee1994building} for 3D inputs. Subsequently, we dilate the skeleton with a diamond kernel of radius $2$ to make it tubular, thereby enlarging the effective area for loss computation around the otherwise thin, single-pixel-wide skeleton. This enhances the stability of the loss by incorporating signals from a greater number of pixels, particularly those in close proximity to the skeleton, which are vital for connectivity. Lastly, for multi-class problems, we multiply the tubed skeleton with the ground truth mask, effectively assigning parts of the skeleton to their respective classes. All of these operations are computationally inexpensive and can be carried out on the CPU during data loading or pre-computed using libraries such as \texttt{scikit-image} \cite{van2014scikit}.

\subsubsection{Soft Recall on Tubed Skeleton}
Following the extraction of our tubed skeleton, we incentivize the network to \textit{include} as much of this skeleton as possible as part of the prediction. This is performed simply by using a \textit{soft recall} loss $\mathcal{L}_{SkelRecall}$ (\cref{eq:skel_recall}), in addition to any existing generic loss $\mathcal{L}_{generic}$ used by the network (for example, Dice Loss, Cross Entropy Loss, \etc).
\begin{equation}
    \mathcal{L}_{SkelRecall} = - \frac{1}{|C|}\sum_{c\in C}\frac{\sum_i Y_{\mathrm{skel},i,c} \cdot \hat{Y}_{i,c}}{\sum_i Y_{\mathrm{skel},i,c}}
    \label{eq:skel_recall}\\
\end{equation}
This vastly improves the connectivity of thin curvilinear structures predicted by a network trained using this loss (\cref{sec:sota_connectiviy}). Additionally, Skeleton Recall Loss is computationally inexpensive in comparison to the use of a differentiable skeletonization, requiring only fractionally more GPU memory and additional time during training (\cref{sec:minimal_overhead}). This facilitates training multi-class segmentation problems as opposed to current differentiable skeleton methods which incur infeasible overheads (\cref{sec:multi_vs_binary}).

\begin{algorithm}[t]
\caption{Tubed Skeletonization}
\label{alg:tubed_skel}
\begin{algorithmic}[1]
\REQUIRE $Y$ are $K$-classed hard targets where $Y_{i,j(,k)} \in [0, K]$
\STATE $Y_\text{bin} \leftarrow Y>0$  \hfill \textit{\% Binarize to foreground and background labels}
\STATE $Y_\text{skel} \leftarrow \text{\textbf{skeletonize}}(Y_\text{bin})$ \hfill \textit{\% Extract binarized skeleton}
\STATE $Y_\text{skel} \leftarrow \text{\textbf{dilate}}(Y_\text{skel})$ \hfill \textit{\% Dilate to create tubed skeleton}
\STATE $Y_\text{mc-skel} \leftarrow Y_\text{skel} \times Y$ \hfill \textit{\% De-binarize to create multi-class tubed skeleton}
\RETURN $Y_\text{mc-skel}$
\end{algorithmic}
\end{algorithm}











\section{Experimental Setup}

\subsection{Datasets}
\label{sec:datasets}

\begin{table}[ht]
    \caption{\textbf{Details of the datasets used for training and evaluation.} Our datasets show wide coverage over a number of thin structure segmentation tasks in natural and medical images. The TopCoW dataset is used both in binary and multi-class settings, in line with the original challenge.}
    \label{tab:datasets}
    \centering
    \begin{tabular}{lcccc}
    \toprule
    \textbf{Dataset} & \textbf{Dims} & \textbf{Type} & \textbf{\# Samples} & \textbf{Target Structure} \\
    \toprule
    Roads\cite{roads} & 2D & binary & 804 & Roads on aerial images\\
    DRIVE\cite{drive} & 2D & binary & 20 & Blood vessels on retina images\\
    Cracks\cite{cracks} & 2D & binary & 5388 & Cracks on concrete structure images\\
    ToothFairy\cite{toothfairycvpr,toothfairydata} & 3D & binary & 138 & Inferior Alveolar Canal\\
    TopCoW\cite{topcow} & 3D & multi-class & 200 & Circle of Willis vessels in the brain\\
    \bottomrule
    \end{tabular}
\end{table}

\noindent We employ five public datasets featuring thin structures for validating the proposed Skeleton Recall Loss. The datasets span natural as well as medical images, covering a range of segmentation challenges, including both binary and multi-class segmentation problems in 2D as well as 3D contexts. An overview of the datasets can be found in \cref{tab:datasets}.
\noindent Among the three 2D datasets used in this study, the Digital Retinal Images for Vessel Extraction (\textbf{DRIVE}) dataset~\cite{drive} was employed, focusing on retinal vessel segmentation. Additionally, structural inspection images designed for concrete crack segmentation (\textbf{Cracks})~\cite{cracks} and aerial images of Massachusetts for road segmentation (\textbf{Roads})~\cite{roads} were included, highlighting the diversity of thin structures in natural and constructed environments. In the 3D domain, we incorporated two cutting-edge medical image segmentation challenge datasets. One of them was \textbf{ToothFairy}\footnote{\url{https://toothfairy.grand-challenge.org/}}, which was a segmentation challenge on 3D Cone-Beam CTs \cite{toothfairycvpr, toothfairydata} featuring the inferior alveolar canal as the target structure. Additionally, the \textbf{TopCoW}\footnote{\url{https://topcow23.grand-challenge.org/}} dataset for topology-aware 3D segmentation of vessels in the Circle of Willis for CTA and MRA data \cite{topcow} was utilized, encompassing binary as well as multi-class segmentation on 13 different subtypes of vessels.
\noindent This diverse set of datasets enables a comprehensive evaluation of the proposed Skeleton Recall Loss, demonstrating generalizability of the method to a wide range of thin structure segmentation challenges in both 2D and 3D contexts.

\subsection{Evaluation Metrics}

We use multiple metrics including overlap, connectivity and topological measures for thorough evaluation of our proposed loss function. An interesting dichotomy of clDice \cite{clDice} is that while it makes for an inefficient loss function for training deep neural networks, it is an effective metric for measuring connectivity. Therefore, following existing guidelines \cite{metricsreloaded} for semantic segmentation of tubular structures, we use clDice \textit{as a metric} in conjunction with Dice similarity coefficient, as our connectivity- and overlap-based measures. Similar to previous work, we also report on topology-based metrics, namely, absolute Betti Number Errors of 0\textsuperscript{th} and 1\textsuperscript{st} Betti Numbers, $\beta_0$ and $\beta_1$. However, in contrast to other work \cite{hu2019topology, clDice}, we calculate the Betti Errors on whole volumes instead of small, randomly extracted patches. Our evaluation strategy is more intuitive in nature and offers better interpretability of the measure, which is especially relevant in medical segmentation tasks.

\subsection{Baseline Loss Functions}
\label{sec:baselines}

We benchmark our proposed loss function against state-of-the-art loss functions targeting thin structure segmentation on the five datasets detailed in \cref{sec:datasets}. Specifically, we compare against: \textbf{1)} \textbf{clDice Loss} \cite{clDice}, the leading method in the field that utilizes approximate differentiable skeletons. \textbf{2)} Additionally, we also compare against a modification of clDice Loss, where we replace the differentiable skeletonization of the original publication by a follow-up of this work~\cite{menten2023skeletonization}. This new method, called \textbf{Topo-clDice Loss} in our evaluations, produces topologically-accurate differentiable skeletons at the cost of even higher computational requirements. 
We note that loss functions based on persistent homologies~\cite{mosinska2019joint, hu2019topology} are excluded from our evaluation which, while related, were surpassed by the clDice Loss \cite{clDice}.

\subsection{Training}

We implement the baseline loss functions (\cref{sec:baselines}), as well as our proposed Skeleton Recall Loss in a powerful medical image segmentation network (nnUNet \cite{Isensee2021}) and a state-of-the-art natural image segmentation network (HRNet~\cite{HRNet}), pretrained on ImageNet~\cite{imagenet}.
We use the examined loss functions for connectivity conservation ($\mathcal{L}_{connectivity}$) in addition to the underlying generic loss ($\mathcal{L}_{generic}$) of our training framework -- a combination of Cross-Entropy and Soft Dice Loss. The connectivity loss is weighted by an additional parameter $w$ as shown in \cref{eq:loss_comparison}.

\begin{equation}
    \mathcal{L} = \mathcal{L}_{generic} + w \cdot \mathcal{L}_{connectivity}
    \label{eq:loss_comparison}
\end{equation}

Our experiments are restricted to two weight configurations $w \in \{0.1, 1.0\}$ in order to curb the influence of extensive hyperparameter tuning. We show a more detailed analysis of the effect of the weight parameter in the Appendix. Additionally, the full set of hyperparameters, optimizers and configurations of nnUNet and HRNet used for training on the different datasets are also provided in the Appendix.

\section{Results and Discussion}

\subsection{Skeleton Recall Loss enables state-of-the-art segmentation of thin structures}
\label{sec:sota_connectiviy}

\begin{table}[!ht]
  \aboverulesep=0ex
  \belowrulesep=0ex
  \caption{\textbf{State-of-the-art segmentation of thin structures.} Quantitative results obtained by incorporating our proposed Skeleton Recall Loss as well as existing thin structure segmentation losses into the loss function of a generic nnUNet backbone. Results are reported on the testset, \textit{except for} DRIVE and Cracks datasets, where we report 5-fold cross validation results due to unavailability of an independent testset.}
  \label{tab:results_thin}
  \centering
  \begin{adjustbox}{width=0.95\textwidth}
  \begin{tabular}{@{\hspace{1mm}}l@{\hspace{1mm}}l@{\hspace{1mm}}|@{\hspace{1mm}}c@{\hspace{1mm}}c@{\hspace{1mm}}|c@{\hspace{1mm}}c@{\hspace{1mm}}}
    \toprule
    \rule{0pt}{1.1EM}
    \textbf{Dataset} & \textbf{Loss configuration} & \textbf{Dice} $\uparrow$ & \textbf{clDice} $\uparrow$ & $\mathbf{\beta_0}$ \textbf{error} $\downarrow$ & $\mathbf{\beta_1}$ \textbf{error} $\downarrow$ \\
    \toprule
    \rule{0pt}{1.1EM}
    \multirow{4.25}{*}{\shortstack{Roads \\ \cite{roads}}} & Default nnUNet & 78.99 & 88.79 & 5.769 & 84.62 \\
     & ~+ clDice Loss & 79.15 & 89.00 & 6.539 & \textbf{82.00} \\
     & ~+ Topo-clDice Loss & 78.94 & 88.62 & 11.00 & 85.92 \\ 
    \cmidrule{2-6}
    \rule{0pt}{1.1EM}
     & ~+ Skeleton Recall Loss \textbf{(\textit{Ours})} & \textbf{79.25} & \textbf{89.06} & \textbf{4.846} & 83.69 \\
    \midrule
    \rule{0pt}{1.1EM}
    \multirow{4.25}{*}{\shortstack{DRIVE \\ \cite{drive}}} & Default nnUNet & 80.87 & 80.26 & 57.00 & 22.80 \\
     & ~+ clDice Loss & \textbf{81.05} & 80.68 & 44.50 & 23.35 \\
     & ~+ Topo-clDice Loss & 80.80 & 80.19 & 46.35 & 23.70 \\ 
    \cmidrule{2-6}
    \rule{0pt}{1.1EM}
     & ~+ Skeleton Recall Loss \textbf{(\textit{Ours})} & 80.99 & \textbf{80.83} & \textbf{38.75} & \textbf{21.50} \\
    \midrule
    \rule{0pt}{1.1EM}
    \multirow{4.25}{*}{\shortstack{Cracks \\ \cite{cracks}}} & Default nnUNet & 94.59 & 95.76 & 0.147 & \textbf{0.0033} \\
     & ~+ clDice Loss & 94.80 & 95.96 & \textbf{0.142} & \textbf{0.0033} \\
     & ~+ Topo-clDice Loss & 94.83 & 96.00 & 0.159 & 0.0035 \\ 
    \cmidrule{2-6}
    \rule{0pt}{1.1EM}
     & ~+ Skeleton Recall Loss \textbf{(\textit{Ours})} & \textbf{94.88} & \textbf{96.04} & 0.148 & 0.0035 \\
    \midrule
    \rule{0pt}{1.1EM}
    \multirow{3.25}{*}{\shortstack{ToothFairy \\ \cite{toothfairycvpr,toothfairydata}}} & Default nnUNet & 71.80 & 89.16 & 0.900 & \textbf{0.0200} \\
     & ~+ clDice Loss & 72.36 & 89.67 & 0.620 & \textbf{0.0200} \\
    \cmidrule{2-6}
    \rule{0pt}{1.1EM}
     & ~+ Skeleton Recall Loss \textbf{(\textit{Ours})} & \textbf{74.42} & \textbf{92.05} & \textbf{0.540} & \textbf{0.0200} \\
    \midrule
    \rule{0pt}{1.1EM}
    \multirow{3.25}{*}{\shortstack{TopCoW \\ binary\\ \cite{topcow}}}  & Default nnUNet & 93.55 & 98.25 & 0.743 & 1.800 \\
     & ~+ clDice Loss & 93.64 & 98.35 & 0.514 & 1.986 \\
    \cmidrule{2-6}
    \rule{0pt}{1.1EM}
     & ~+ Skeleton Recall Loss \textbf{(\textit{Ours})} & \textbf{93.72} & \textbf{98.48} & \textbf{0.500} & \textbf{1.586} \\
    \midrule
    \rule{0pt}{1.1EM}
    \multirow{3.25}{*}{\shortstack{TopCoW\\multi-class}} & Default nnUNet & 85.36 & 93.68 & \textbf{0.137} & 0.0571 \\
     & ~+ clDice Loss & \multicolumn{4}{c}{-- \textit{Out Of Memory} --} \\
    \cmidrule{2-6}
    \rule{0pt}{1.1EM}
     & ~+ Skeleton Recall Loss \textbf{(\textit{Ours})} & \textbf{86.59} & \textbf{94.35} & 0.151 & \textbf{0.056} \\
  \bottomrule
  \end{tabular}
  \end{adjustbox}
\end{table}

\begin{figure}[!ht]
     \centering
     \begin{subfigure}[b]{0.196\textwidth}
         \caption*{\textbf{Image}}
         \centering
         \includegraphics[width=\textwidth]{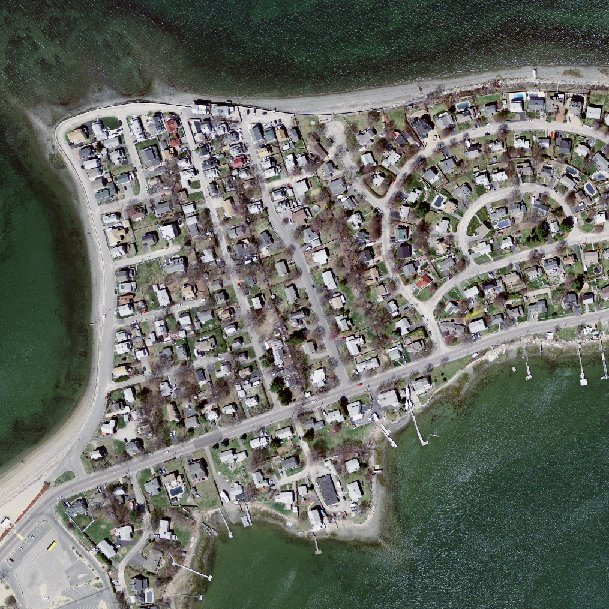}
     \end{subfigure}%
     \hfill
     \begin{subfigure}[b]{0.196\textwidth}
         \caption*{\textbf{Ground Truth}}
         \centering
         \includegraphics[width=\textwidth]{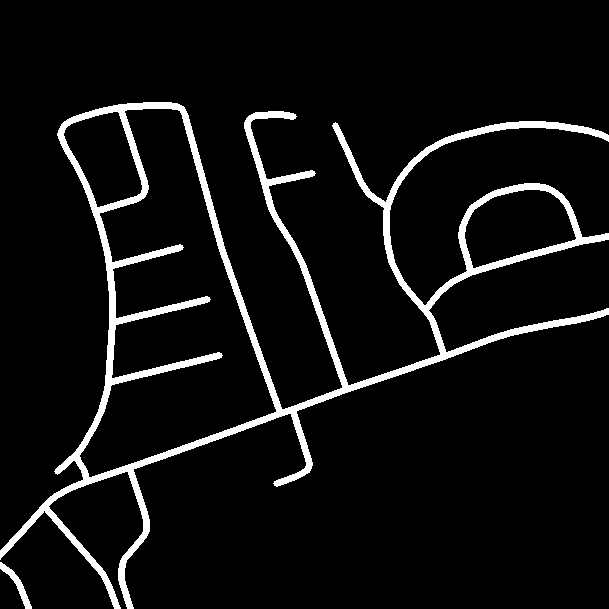}
     \end{subfigure}%
     \hfill
     \begin{subfigure}[b]{0.196\textwidth}
         \caption*{\textbf{nnUNet}}
         \centering
         \includegraphics[width=\textwidth]{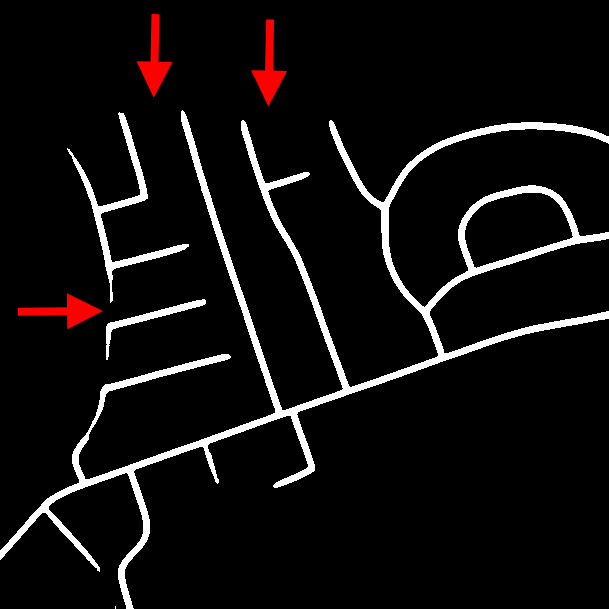}
     \end{subfigure}%
     \hfill
     \begin{subfigure}[b]{0.196\textwidth}
         \caption*{\textbf{+ clDice Loss}}
         \centering
         \includegraphics[width=\textwidth]{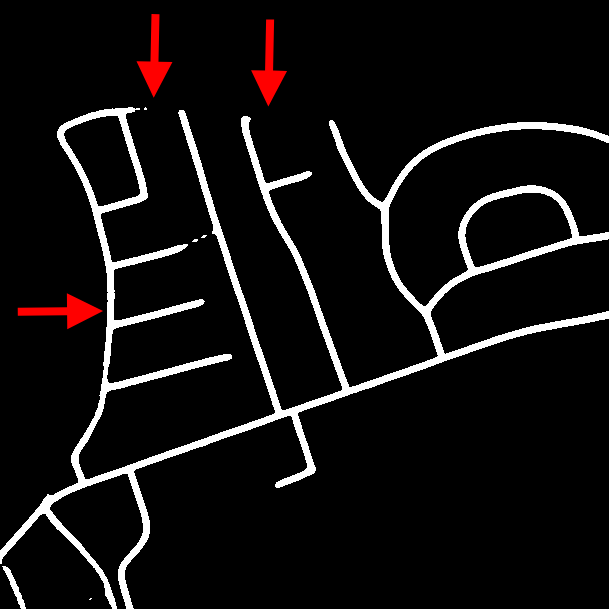}
     \end{subfigure}%
     \hfill
     \begin{subfigure}[b]{0.196\textwidth}
         \caption*{\textbf{+ \textit{Ours}}}
         \centering
         \includegraphics[width=\textwidth]{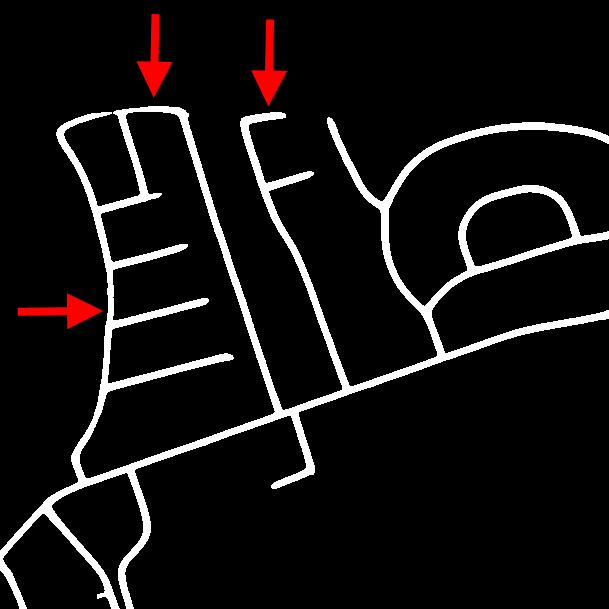}
     \end{subfigure}%
     \newline
     \begin{subfigure}[b]{0.196\textwidth}
         \centering
         \includegraphics[width=\textwidth]{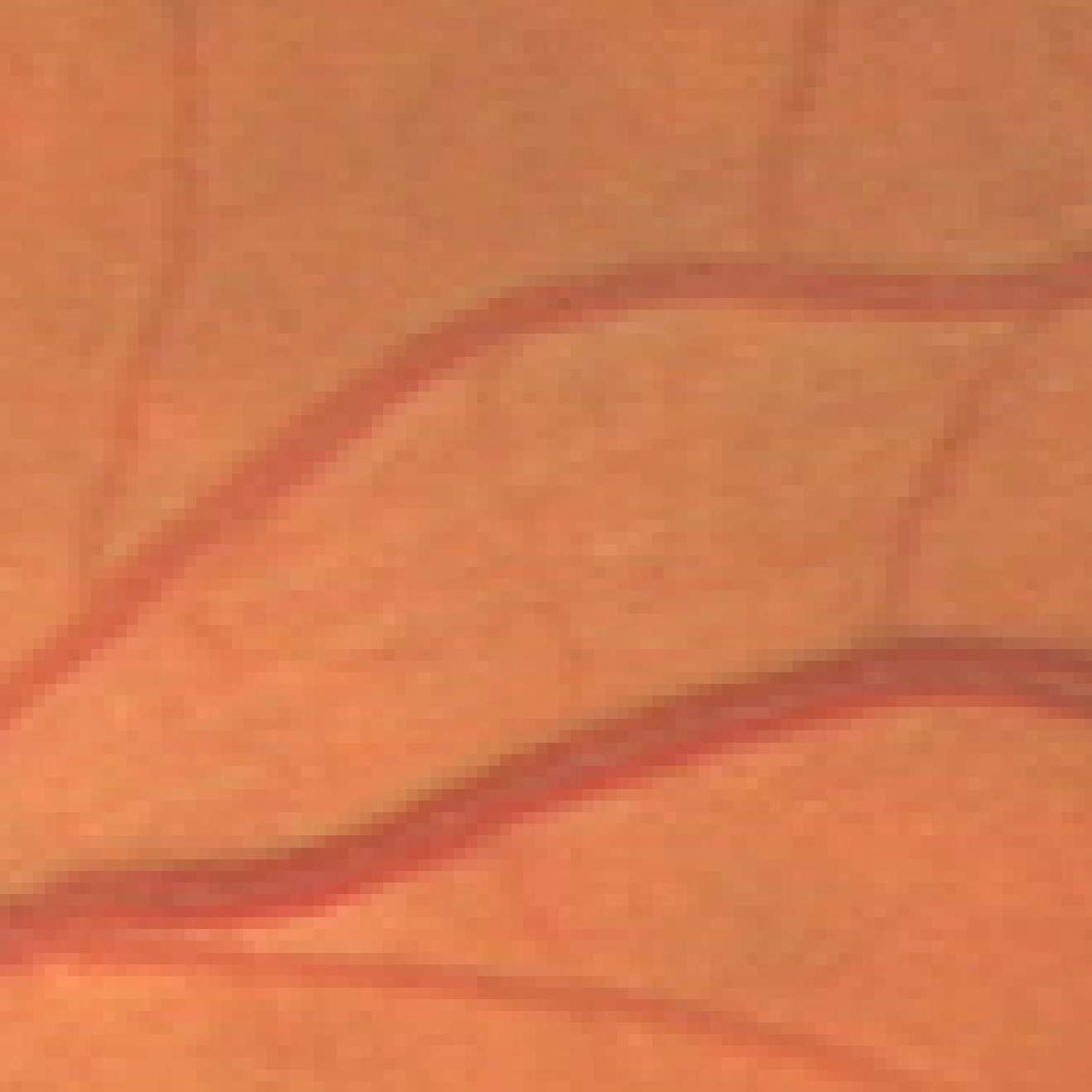}
     \end{subfigure}%
     \hfill
     \begin{subfigure}[b]{0.196\textwidth}
         \centering
         \includegraphics[width=\textwidth]{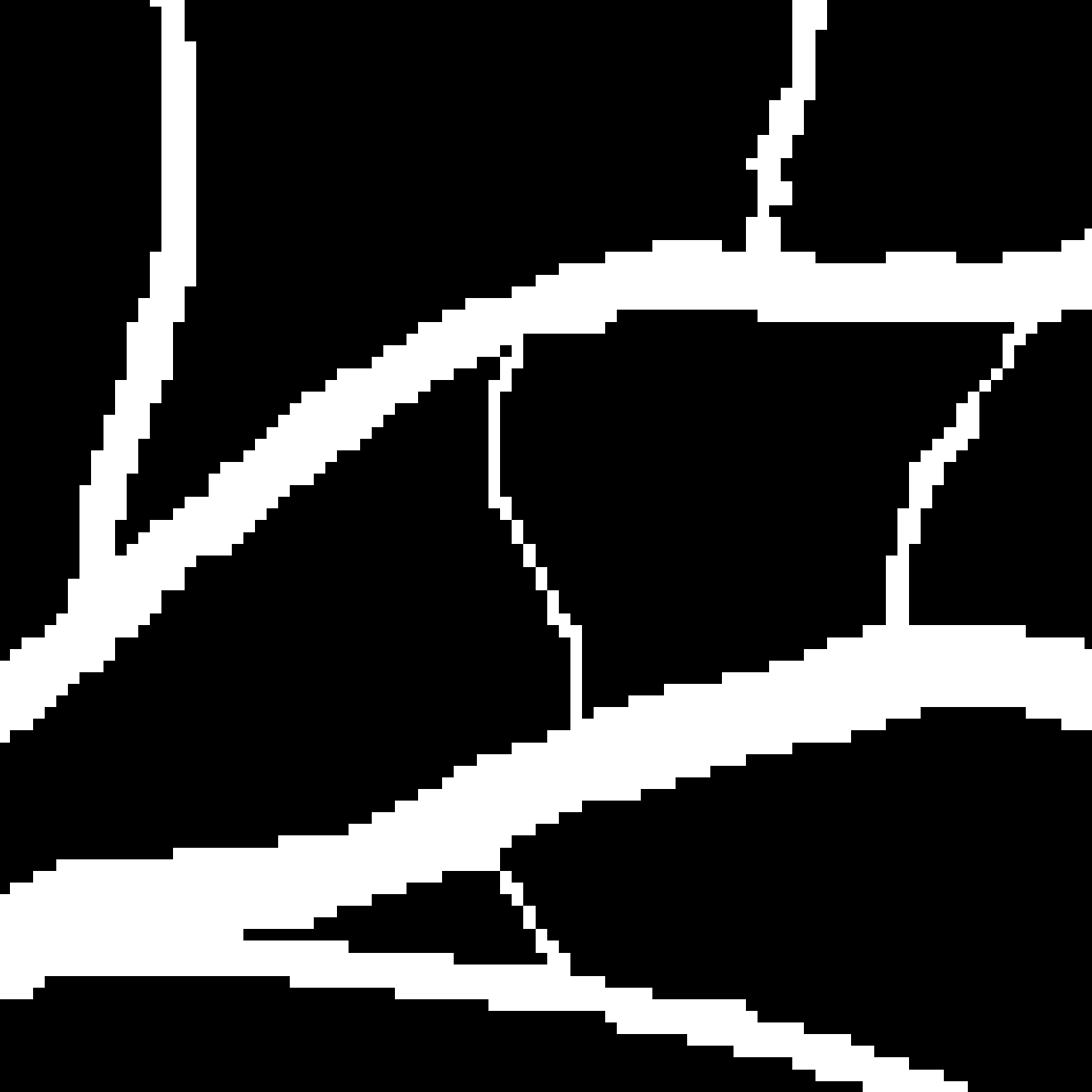}
     \end{subfigure}%
     \hfill
     \begin{subfigure}[b]{0.196\textwidth}
         \centering
         \includegraphics[width=\textwidth]{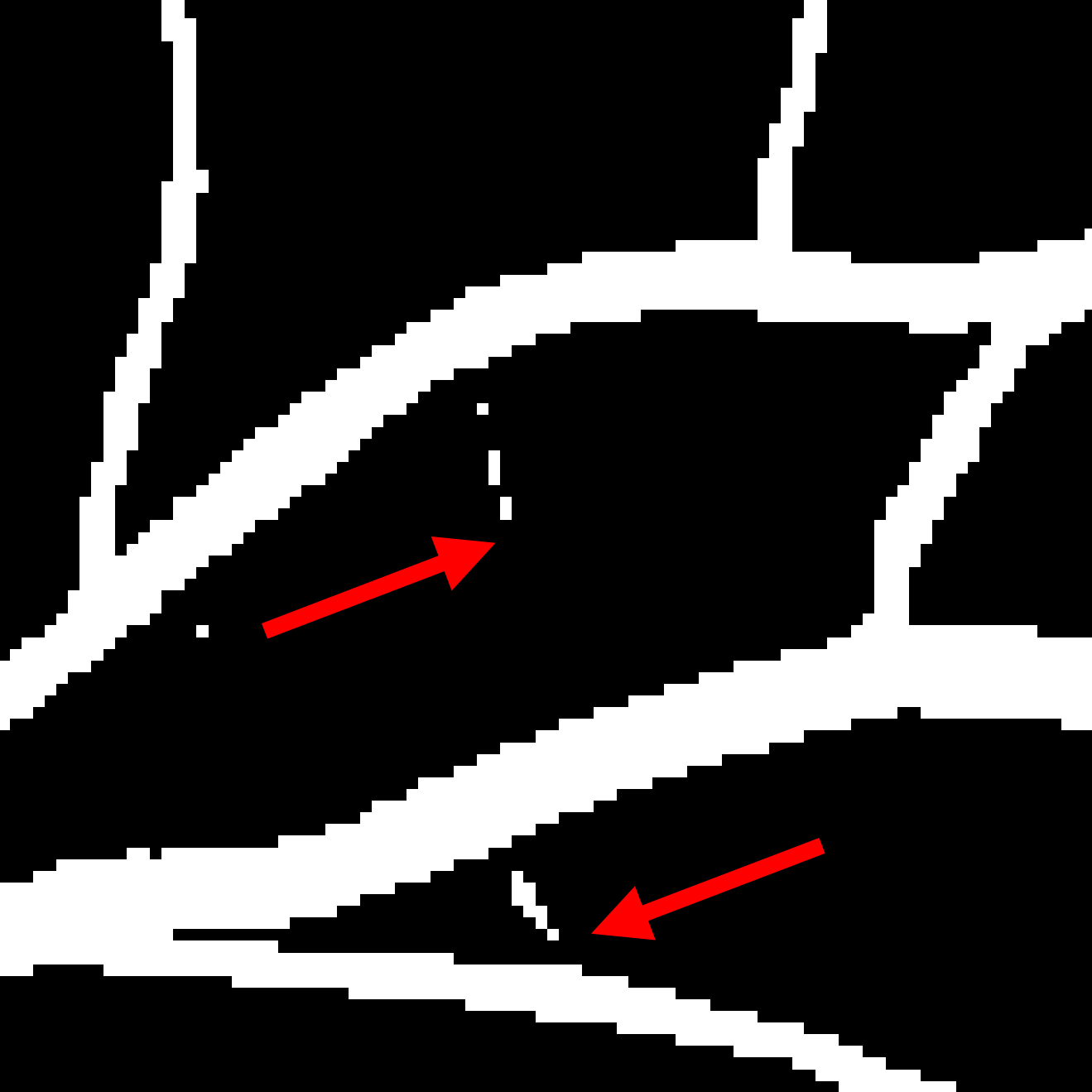}
     \end{subfigure}%
     \hfill
     \begin{subfigure}[b]{0.196\textwidth}
         \centering
         \includegraphics[width=\textwidth]{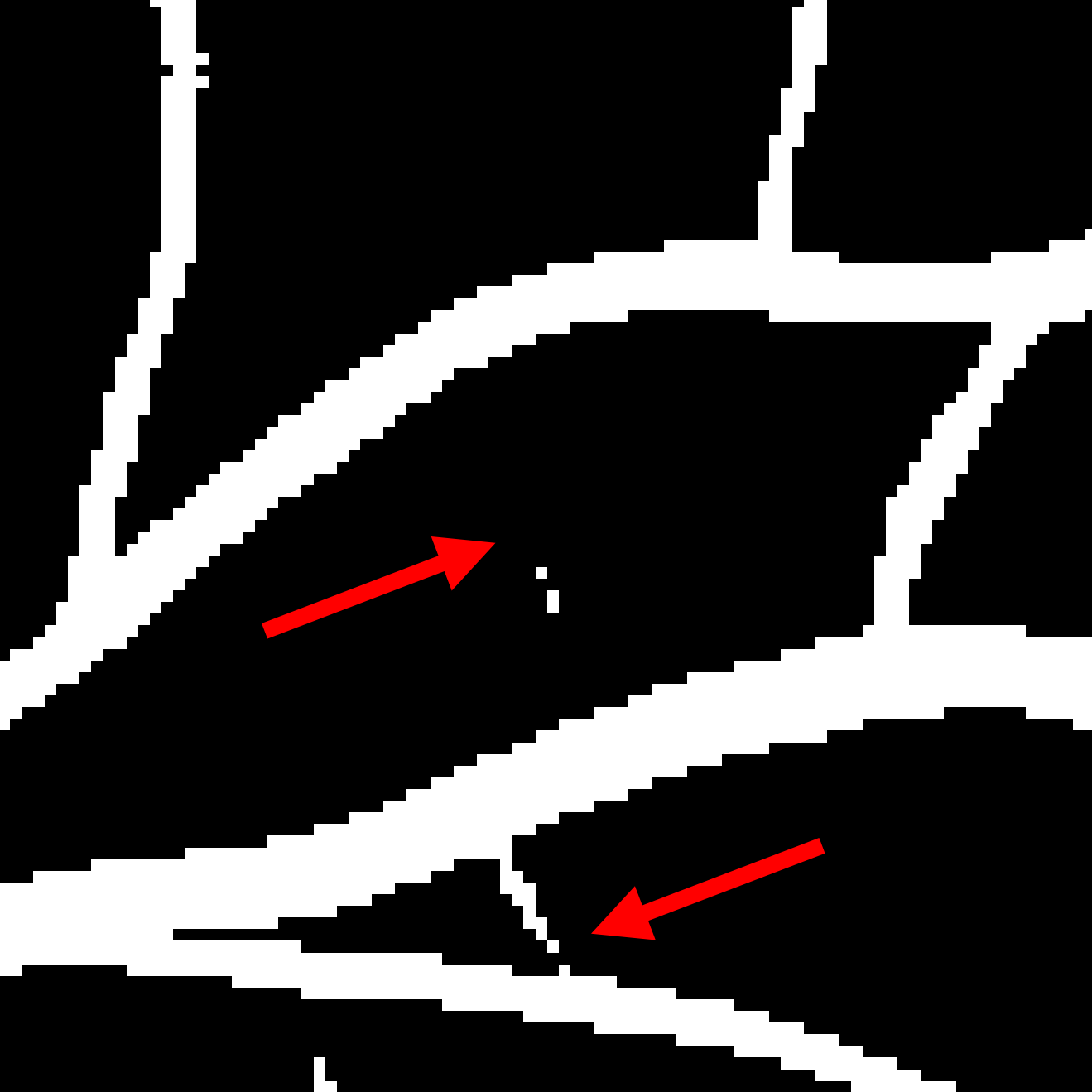}
     \end{subfigure}%
     \hfill
     \begin{subfigure}[b]{0.196\textwidth}
         \centering
         \includegraphics[width=\textwidth]{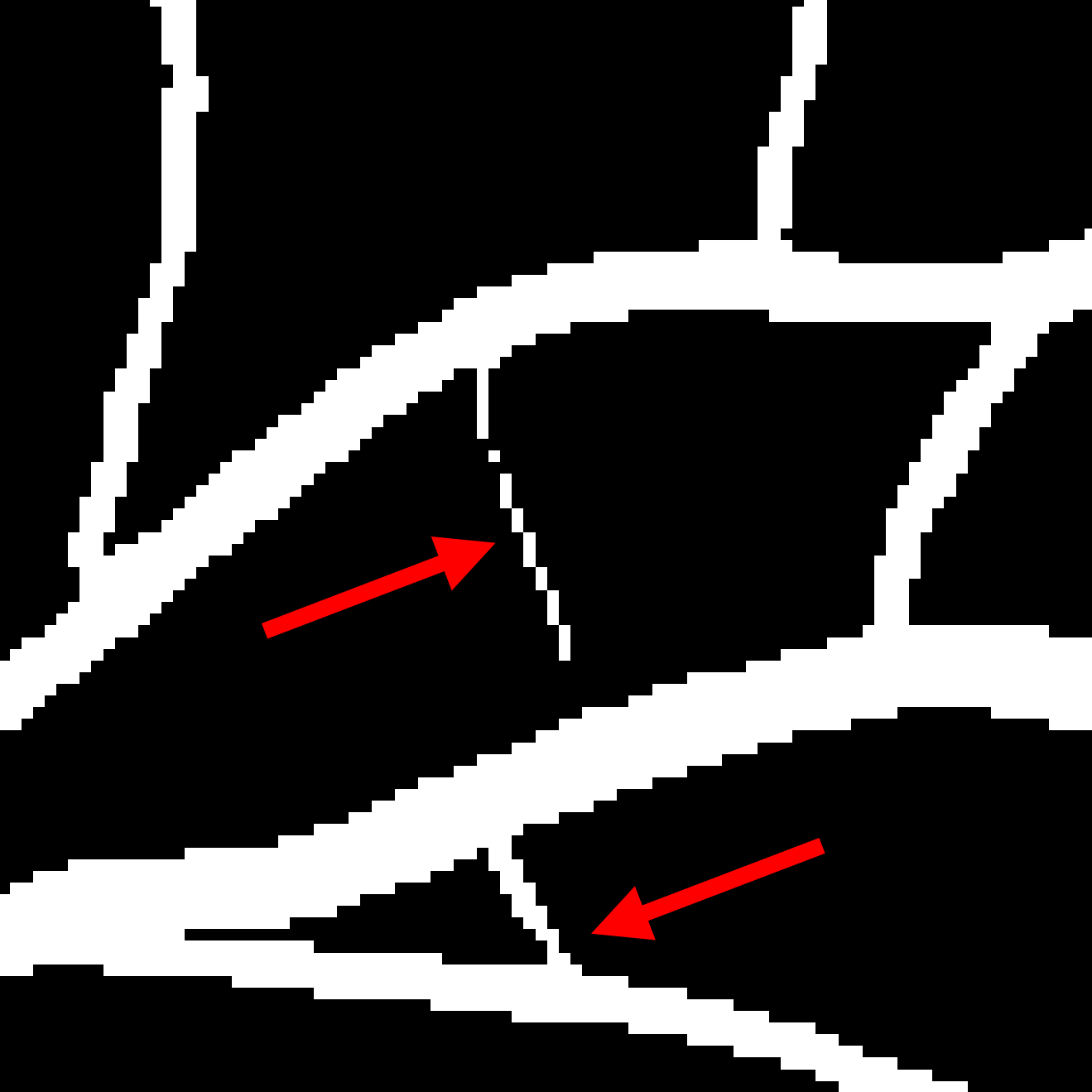}
     \end{subfigure}%
     \newline
     \begin{subfigure}[b]{0.196\textwidth}
         \centering
         \includegraphics[width=\textwidth]{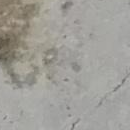}
     \end{subfigure}%
     \hfill
     \begin{subfigure}[b]{0.196\textwidth}
         \centering
         \includegraphics[width=\textwidth]{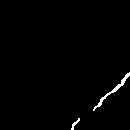}
     \end{subfigure}%
     \hfill
     \begin{subfigure}[b]{0.196\textwidth}
         \centering
         \includegraphics[width=\textwidth]{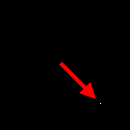}
     \end{subfigure}%
     \hfill
     \begin{subfigure}[b]{0.196\textwidth}
         \centering
         \includegraphics[width=\textwidth]{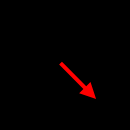}
     \end{subfigure}%
     \hfill
     \begin{subfigure}[b]{0.196\textwidth}
         \centering
         \includegraphics[width=\textwidth]{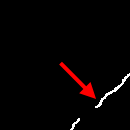}
     \end{subfigure}%
     \newline
     \begin{subfigure}[b]{0.196\textwidth}
         \centering
         \includegraphics[width=\textwidth]{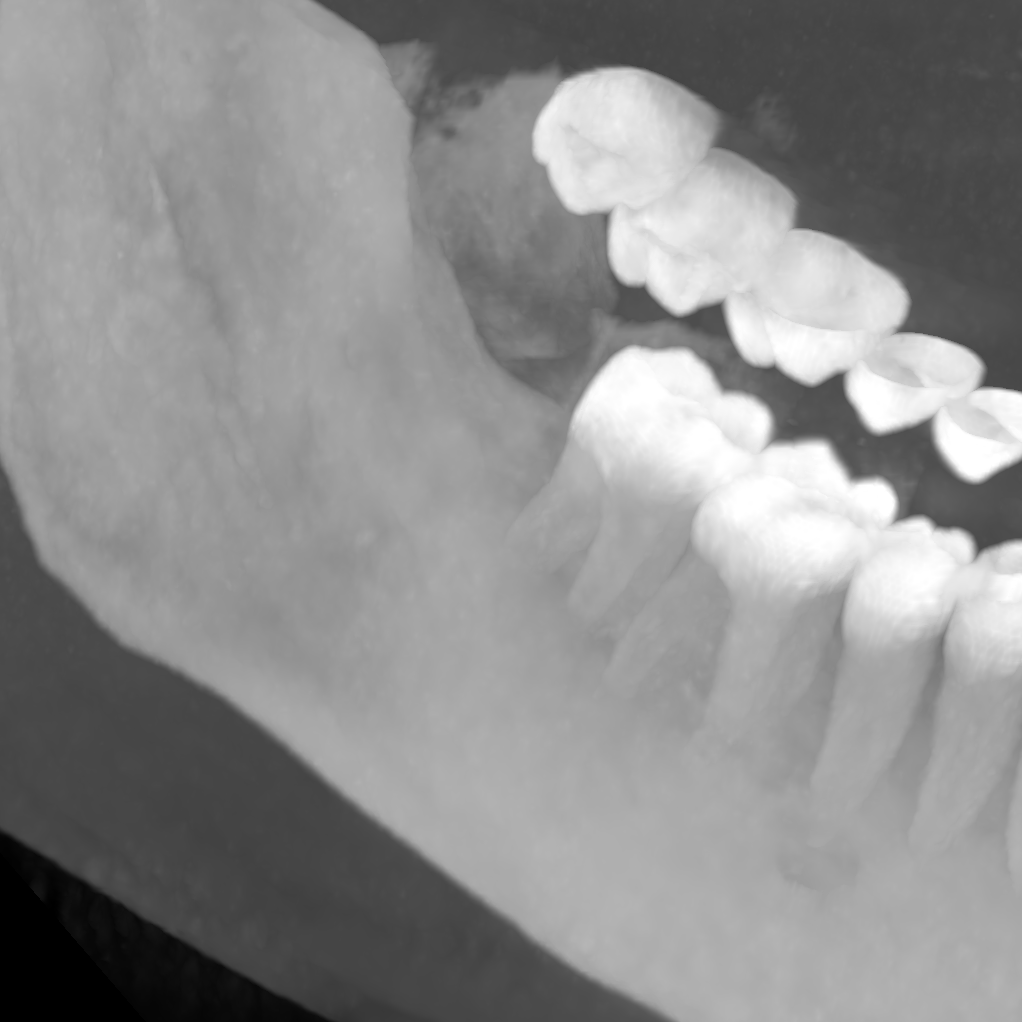}
     \end{subfigure}%
     \hfill
     \begin{subfigure}[b]{0.196\textwidth}
         \centering
         \includegraphics[width=\textwidth]{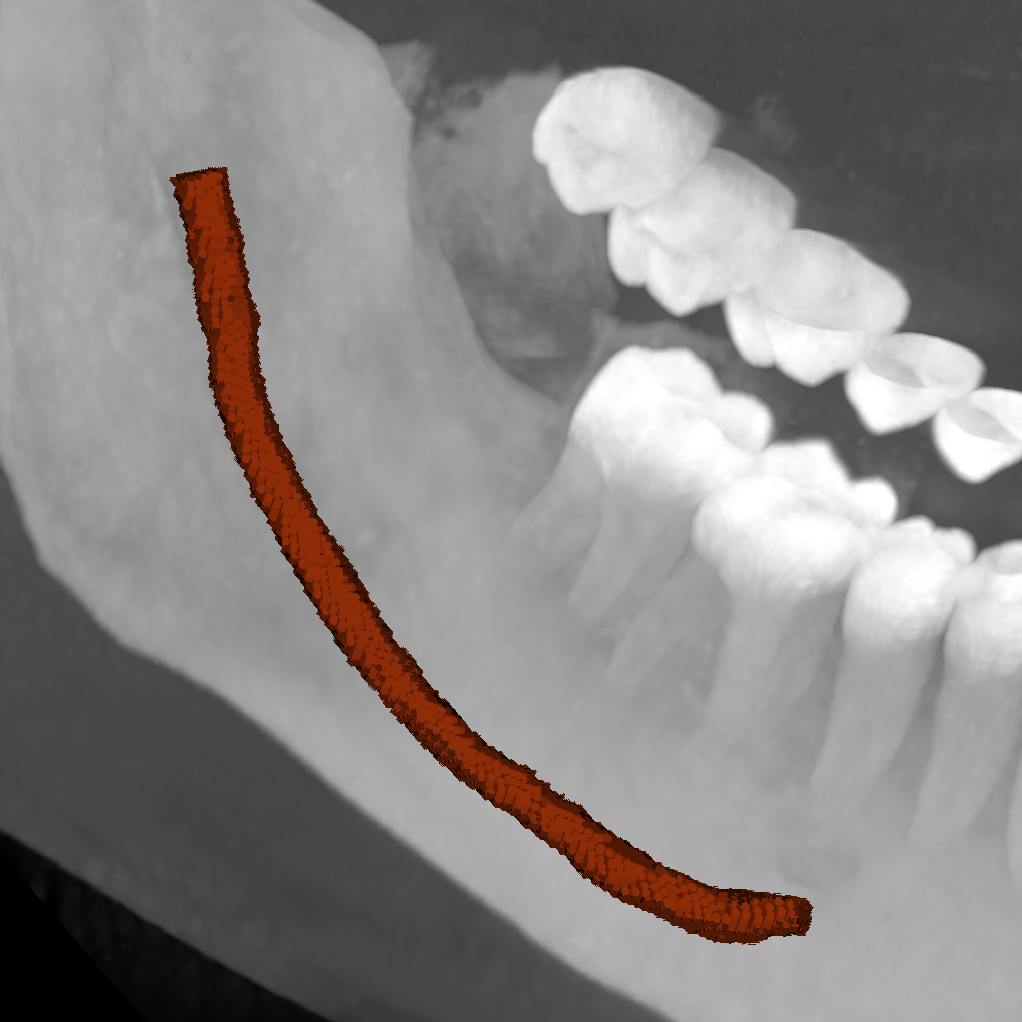}
     \end{subfigure}%
     \hfill
     \begin{subfigure}[b]{0.196\textwidth}
         \centering
         \includegraphics[width=\textwidth]{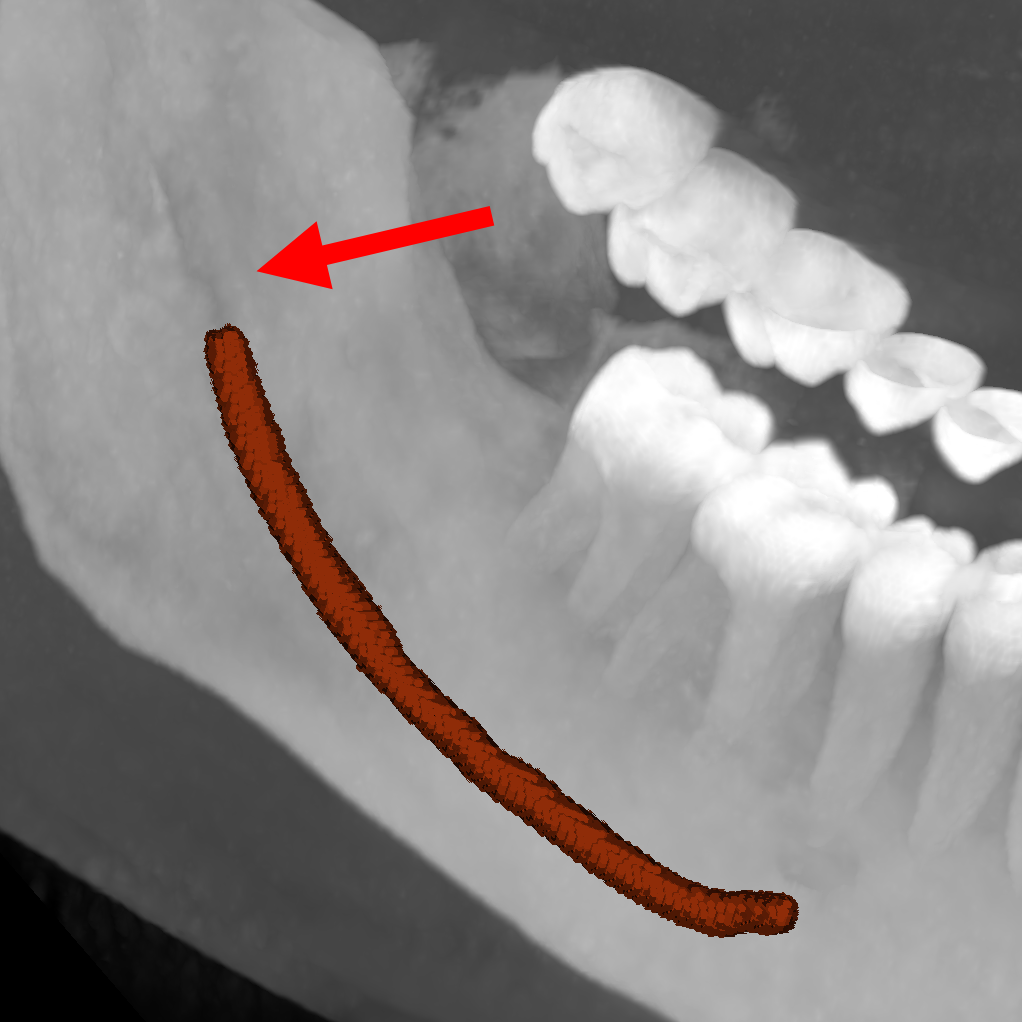}
     \end{subfigure}%
     \hfill
     \begin{subfigure}[b]{0.196\textwidth}
         \centering
         \includegraphics[width=\textwidth]{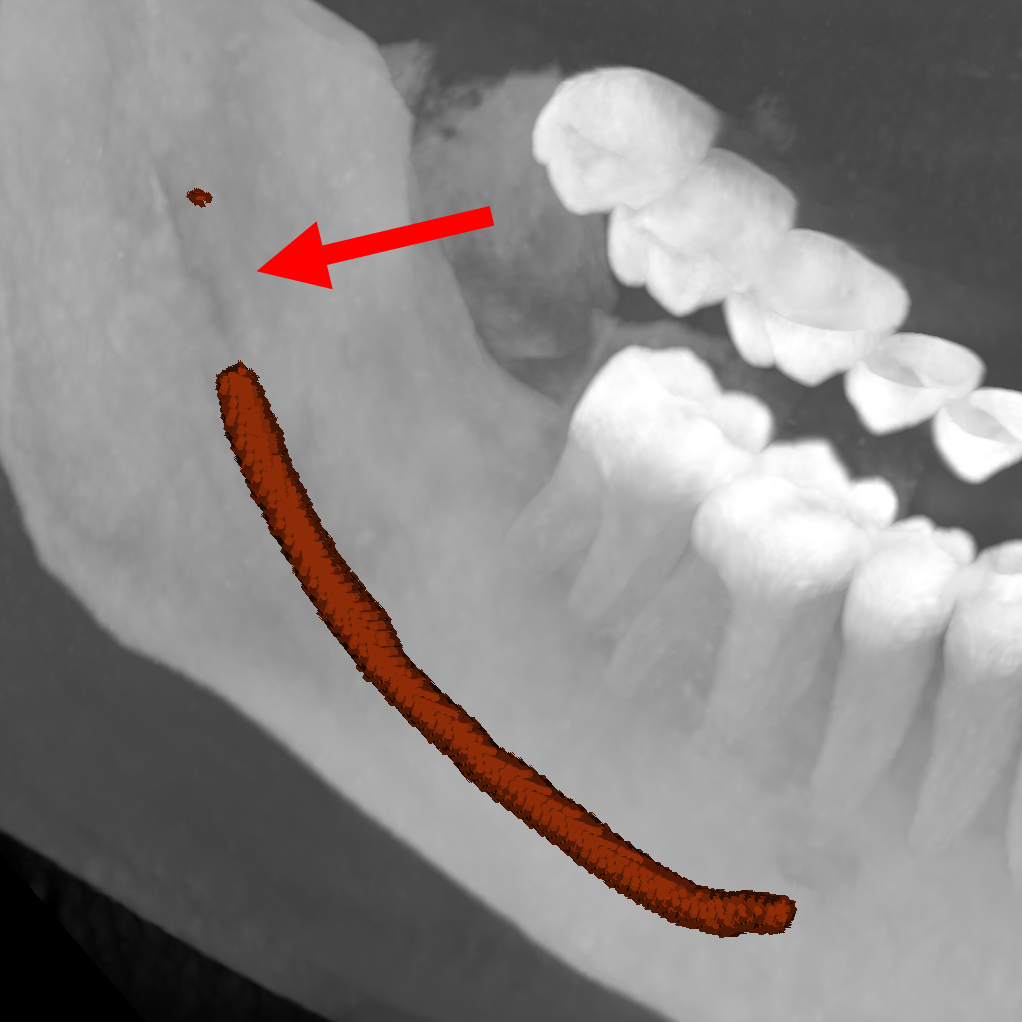}
     \end{subfigure}%
     \hfill
     \begin{subfigure}[b]{0.196\textwidth}
         \centering
         \includegraphics[width=\textwidth]{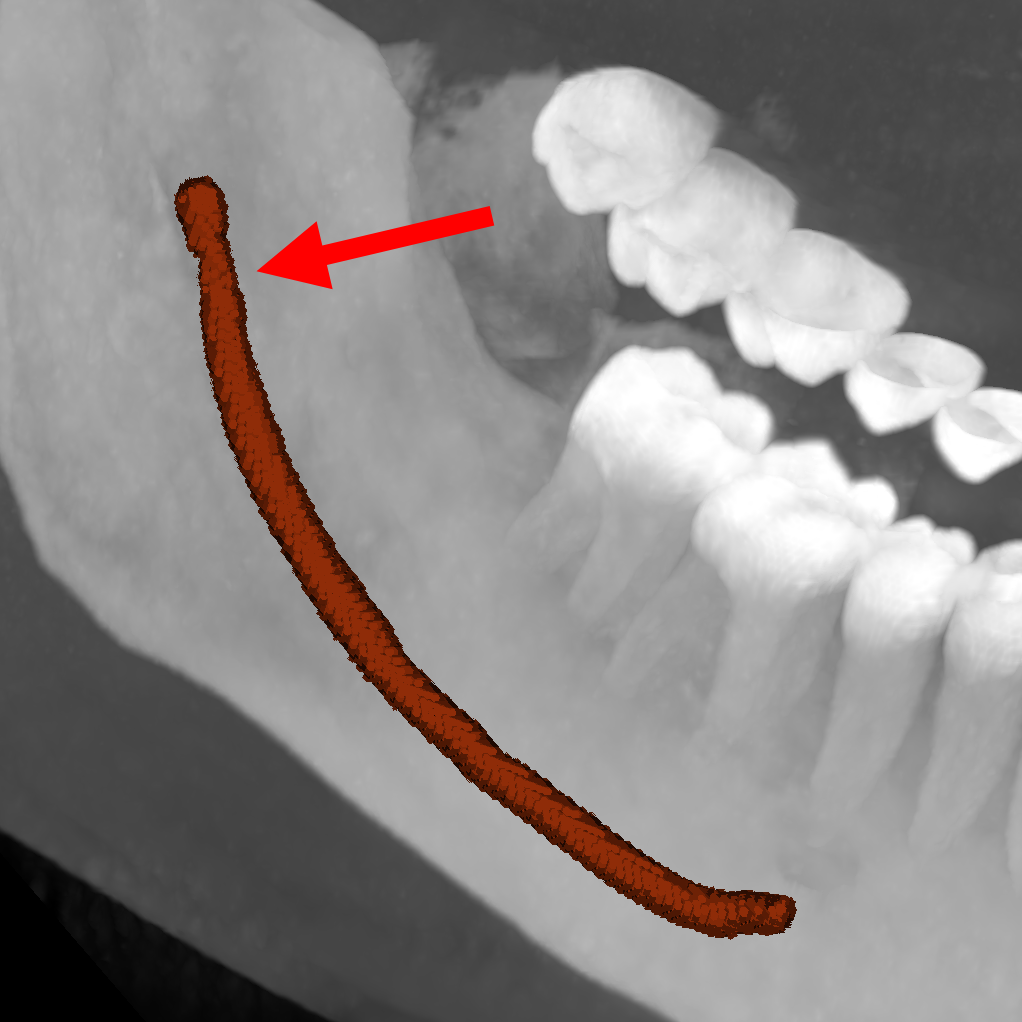}
     \end{subfigure}%
     \newline
     \begin{subfigure}[b]{0.196\textwidth}
         \centering
         \includegraphics[width=\textwidth]{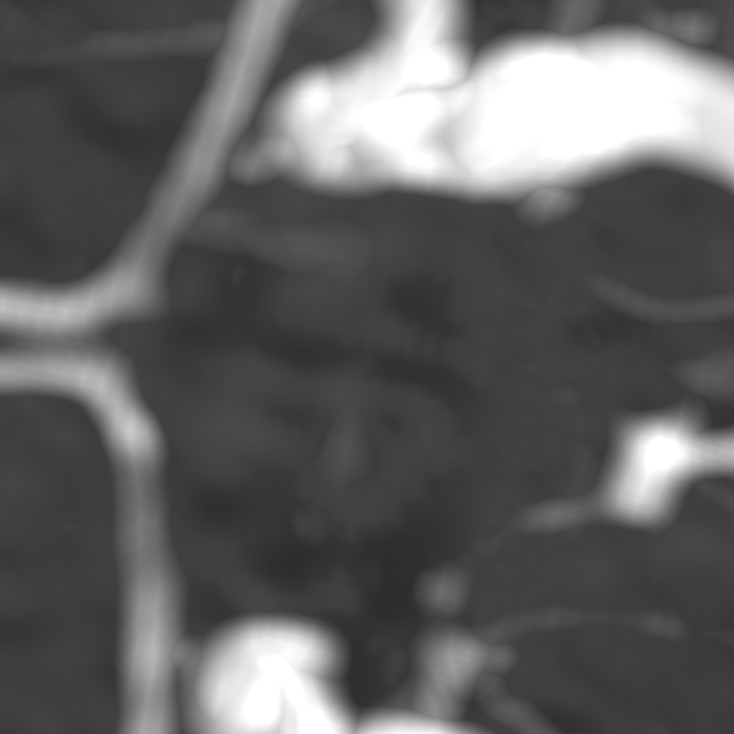}
     \end{subfigure}%
     \hfill
     \begin{subfigure}[b]{0.196\textwidth}
         \centering
         \includegraphics[width=\textwidth]{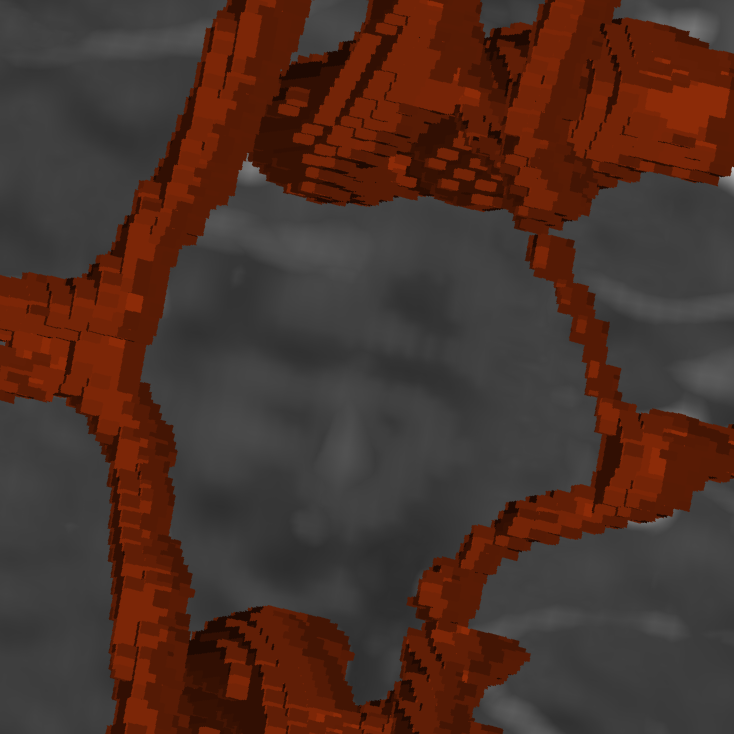}
     \end{subfigure}%
     \hfill
     \begin{subfigure}[b]{0.196\textwidth}
         \centering
         \includegraphics[width=\textwidth]{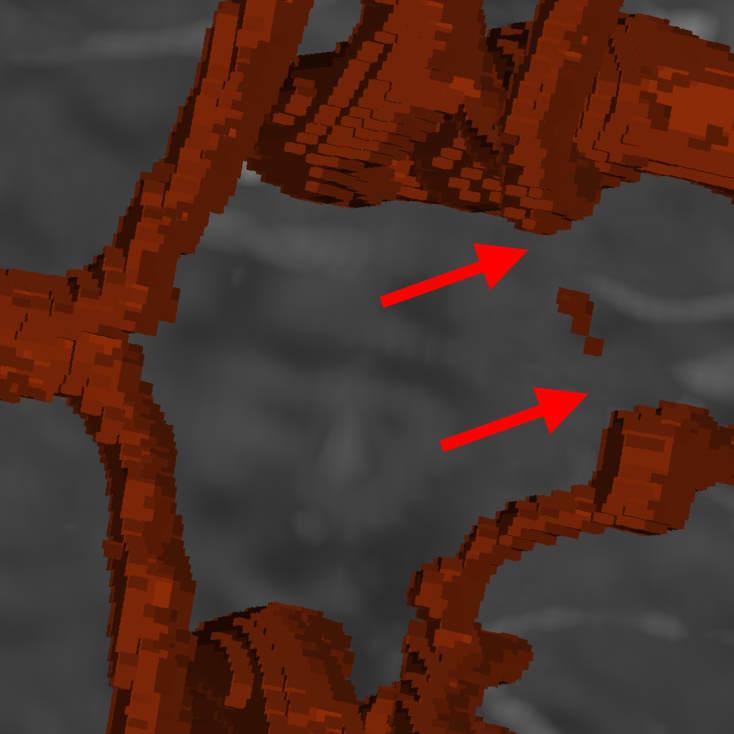}
     \end{subfigure}%
     \hfill
     \begin{subfigure}[b]{0.196\textwidth}
         \centering
         \includegraphics[width=\textwidth]{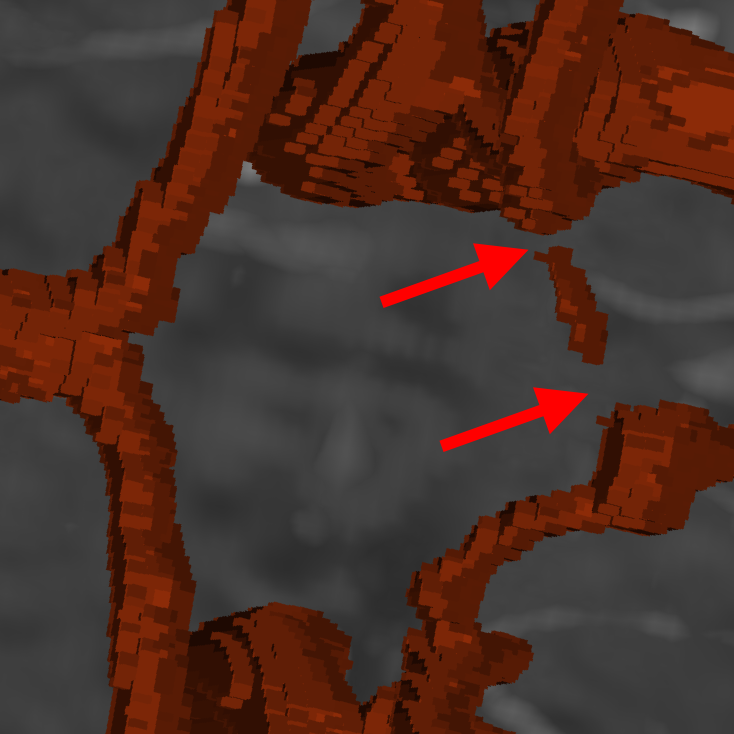}
     \end{subfigure}%
     \hfill
     \begin{subfigure}[b]{0.196\textwidth}
         \centering
         \includegraphics[width=\textwidth]{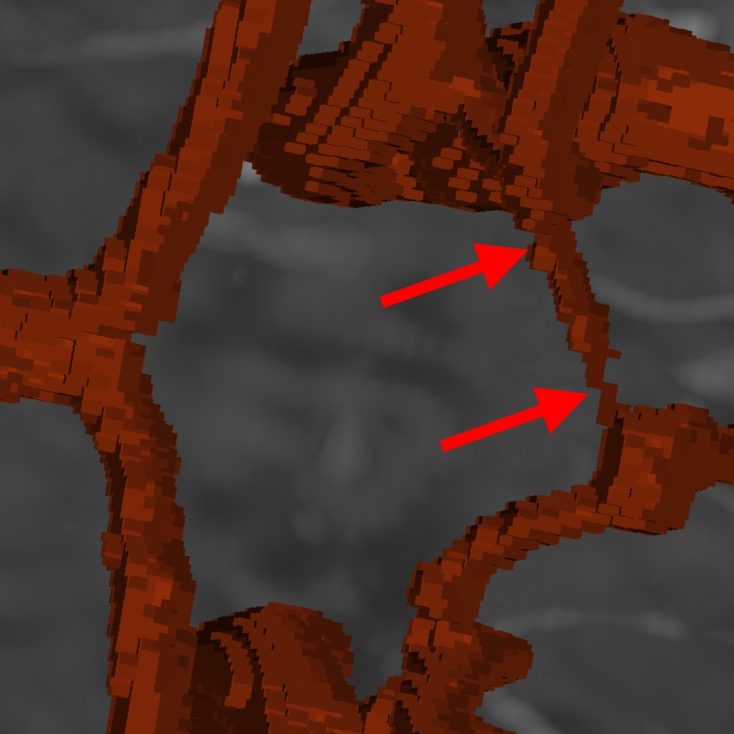}
     \end{subfigure}%
    \caption{\textbf{Connectivity conservation in qualitative results on 5 datasets.} nnUNet with conventional segmentation losses performs well in adequately delineating general structures, particularly thicker ones. However, challenges arise in accurately capturing thin structures and maintaining connectivity within the segmentation. This is demonstrated on examples from (top to bottom) \textbf{Roads}, \textbf{DRIVE}, \textbf{Cracks}, \textbf{Toothfairy} and \textbf{TopCoW} datasets. Augmenting the model with clDice Loss yields some improvement but falls short in addressing connectivity issues. In contrast, our proposed Skeleton Recall Loss demonstrates enhanced preservation of topology and improved connectivity in segmentation outputs.}
    \label{fig:qualitative}
\end{figure}

The obtained results in \cref{tab:results_thin} clearly show that our proposed Skeleton Recall Loss consistently surpasses previous thin structure segmentation losses on almost all datasets. For concrete crack segmentation\cite{cracks}, the results indicate better Dice and clDice performance at the cost of slightly worse Betti numbers than clDice Loss. However, Skeleton Recall Loss demonstrates the best clDice and Betti numbers for retinal vessel segmentation\cite{drive}, yielding a Dice score just marginally behind clDice Loss. Notably, for the three datasets with an independent testset available, specifically Roads\cite{roads} and both of the 3D datasets, ToothFairy\cite{toothfairycvpr,toothfairydata} and TopCoW\cite{topcow}, we observe superior performance of our proposed Skeleton Recall Loss. This is further demonstrated by the qualitative results given in \cref{fig:qualitative}. Skeleton Recall Loss is also seen to be better than baselines, on both binary as well as multi-class settings of TopCoW as elaborated in following sections. We obtain this state-of-the-art performance while being architecture agnostic (\cref{sec:arch_agnostic}) as well as overwhelmingly resource efficient (\cref{sec:minimal_overhead}).

\subsection{Skeleton Recall Loss is architecture agnostic}
\label{sec:arch_agnostic}
While Skeleton Recall Loss demonstrates state-of-the-art performance using\linebreak \textit{nnUNet} as a backbone framework in \cref{tab:results_thin}, it is not restricted to specialized architectures. We highlight this in \cref{tab:results_hrnet} where \textit{HRNet} \cite{HRNet}, a state-of-the-art 2D architecture for natural image segmentation, is used as the backbone. This leads to similar benefits on connectivity conservation using Skeleton Recall Loss during training on our 2D datasets. Skeleton Recall Loss is seen to exceed the connectivity conserving performance (as demonstrated by the clDice metric) on 2 out of 3 datasets, while being comparable on the remaining one. Our overall superiority over all metrics demonstrates that Skeleton Recall Loss is architecture agnostic and can be used as a loss in training arbitrary deep architectures for connectivity-conserving segmentation of thin structures.

\begin{table}[!htbp]
  \aboverulesep=0ex
  \belowrulesep=0ex
  \caption{\textbf{Skeleton Recall Loss is architecture agnostic.} Quantitative results using HRNet, a state-of-the-art 2D network, on all examined 2D datasets. Skeleton Recall Loss demonstrates accurate segmentation including effective connectivity conservation, without explicit reliance on a particular deep neural network architecture.}
  \label{tab:results_hrnet}
  \centering
  \begin{adjustbox}{width=0.95\textwidth}
  \begin{tabular}{@{\hspace{1mm}}l@{\hspace{1mm}}l@{\hspace{1mm}}|@{\hspace{1mm}}c@{\hspace{1mm}}c@{\hspace{1mm}}|c@{\hspace{1mm}}c@{\hspace{1mm}}}
    \toprule
    \rule{0pt}{1.1EM}
    \textbf{Dataset} & \textbf{Loss configuration} & \textbf{Dice} $\uparrow$ & \textbf{clDice} $\uparrow$ & $\mathbf{\beta_0}$ \textbf{error} $\downarrow$ & $\mathbf{\beta_1}$ \textbf{error} $\downarrow$ \\
    \toprule
    \rule{0pt}{1.1EM}
    \multirow{3}{*}{\shortstack{Roads}} & HRNet & 77.91 & 87.63 & 18.08 & 83.15 \\
     & ~+ clDice Loss & 78.05 & \textbf{88.21} & \textbf{8.846} & 82.31 \\ 
    \cmidrule{2-6}
    \rule{0pt}{1.1EM}
     & ~+ Skeleton Recall Loss \textit{(\textbf{Ours})} & \textbf{78.25} & 88.14 & 15.38 & \textbf{81.31} \\
    \midrule
    \rule{0pt}{1.1EM}
    \multirow{3}{*}{\shortstack{DRIVE}} & HRNet & 80.34 & 77.83 & 149.3 & 51.25 \\
     & ~+ clDice Loss & \textbf{80.63} & 81.65 & 65.75 & 46.25 \\ 
    \cmidrule{2-6}
    \rule{0pt}{1.1EM}
     & ~+ Skeleton Recall Loss \textit{(\textbf{Ours})} & 76.08 & \textbf{84.20} & \textbf{27.25} & \textbf{38.50} \\
    \midrule
    \rule{0pt}{1.1EM}
    \multirow{3}{*}{\shortstack{Cracks}} & HRNet & 94.09 & 95.39 & 0.1985 & 0.0037 \\
     & ~+ clDice Loss & 94.67 & 96.01 & \textbf{0.1558} & 0.0037 \\ 
    \cmidrule{2-6}
    \rule{0pt}{1.1EM}
     & ~+ Skeleton Recall Loss \textit{(\textbf{Ours})} & \textbf{95.03} & \textbf{96.26} & 0.1596 & \textbf{0.0019} \\
  \bottomrule
  \end{tabular}
  \end{adjustbox}
\end{table}

\subsection{Connectivity conservation with minimal overheads}
\label{sec:minimal_overhead}

\begin{figure}[b!]
    \centering
    \includegraphics[width=0.49\textwidth]{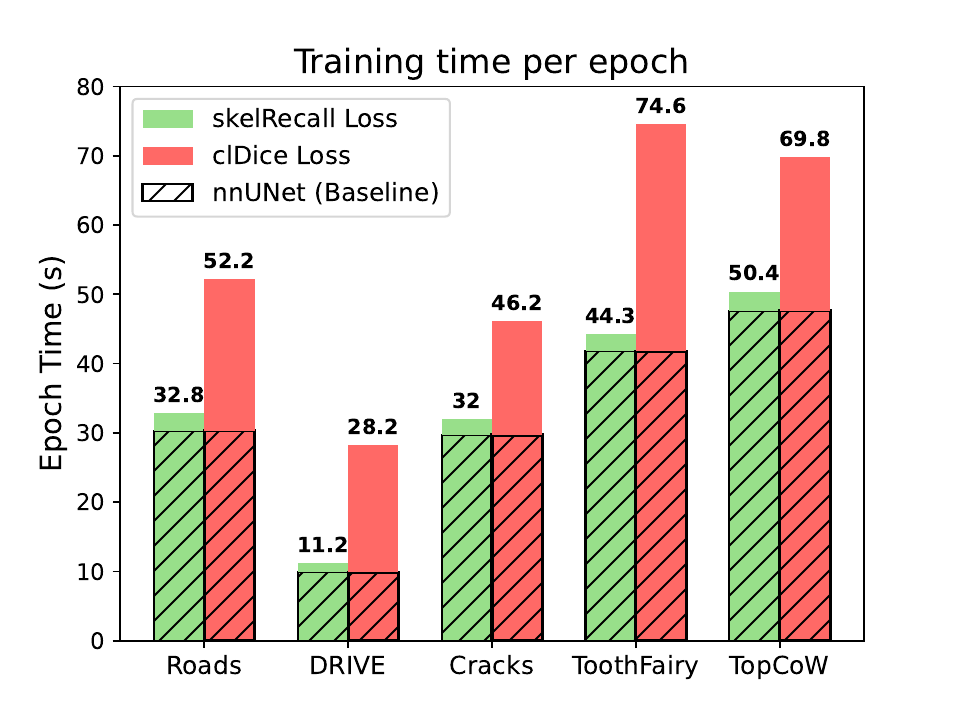}
    \includegraphics[width=0.49\textwidth]{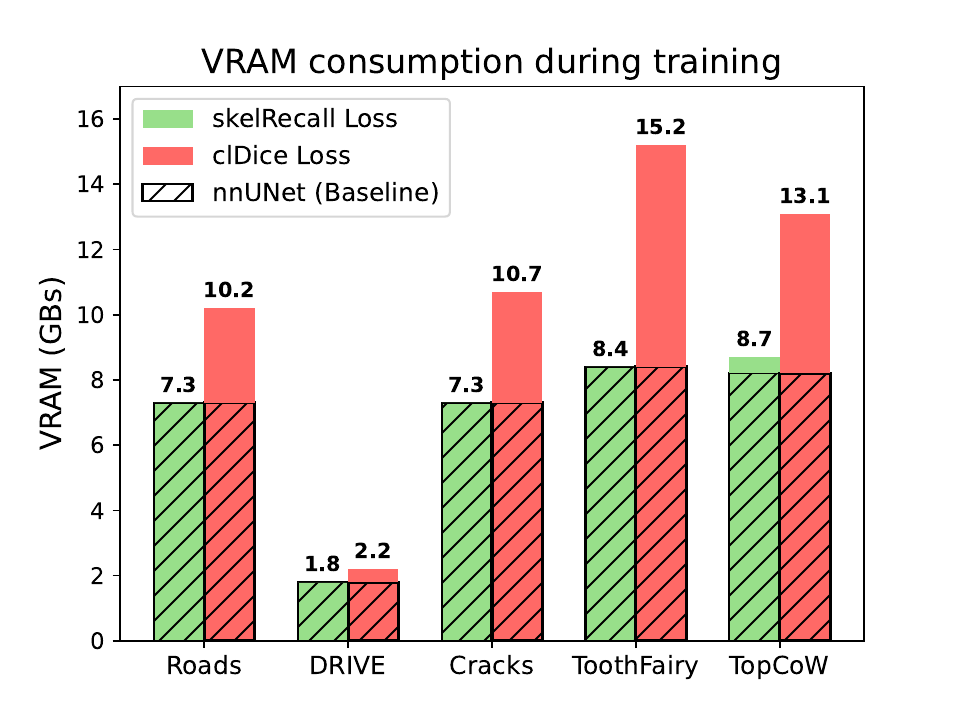}
    \caption{\textbf{Efficient Resource Utilization for Binary Segmentation.} The figures depict the additional training time per epoch and memory requirements caused by employing \textit{Skeleton Recall Loss} and clDice Loss compared to the standard network training (nnUnet, dashed lines) across all assessed datasets. While Skeleton Recall Loss shows minimal increase in VRAM usage and negligible changes in epoch duration, clDice Loss introduces notable overhead in both time and memory. For example, clDice Loss more than doubles the epoch duration for DRIVE or almost doubles VRAM usage for Toothfairy.}
    \label{fig:bin_mem_time_plots}
\end{figure}

\subsubsection{Efficient binary segmentation of thin structures}
\label{sec:memory_util_binary}
A plurality of tasks in the segmentation of thin curvilinear structures have historically been binary in nature. As competing state-of-the-art differentiable skeleton methods were developed for the binary scenario, we consider this to be where such methods should also be most competitive. However, Skeleton Recall Loss does not only provide state-of-the-art connectivity-conserving thin structure segmentation performance, as seen in \cref{sec:sota_connectiviy} and \cref{sec:arch_agnostic}, but it can do so while using \textit{only fractional GPU memory and training time} compared to existing methods, as shown in \cref{fig:bin_mem_time_plots}. 
Differentiable skeleton based methods require a GPU-based skeleton computation \cite{clDice} or prediction \cite{rouge2023cascaded}. For our differentiable skeleton baseline clDice Loss, this leads to approximately $88\%$ additional training time and $52\%$ more VRAM consumption compared to the plain nnUNet backbone when averaged across our 5 datasets (excluding multi-class TopCoW). Remarkably, our method Skeleton Recall Loss does the same at \textit{only an additional $\mathbf{8\%}$ training time and $\mathbf{2\%}$ higher VRAM consumption}. 
This illustrates that Skeleton Recall Loss categorically outperforms traditional differentiable skeleton-based methods on binary settings, which they were developed for, in terms of resource efficiency.

\subsubsection{Enabling multi-class segmentation of thin structures}
\label{sec:multi_vs_binary}

Binary segmentation has historically sufficed for many image analysis tasks across various domains. However, as the demand for finer-grained analysis grows, transitioning to multi-class segmentation becomes increasingly vital. This shift is especially pertinent in medical contexts due to the prevalence of thin structures where binary segmentation may not adequately capture the complexity of anatomical features. For instance, the recent TopCoW challenge\cite{topcow} revealed that binary segmentation of brain vessels can be deemed as sufficiently solved, approaching inter-rater agreement in the Dice score. However, differentiating between the different vessels still remains a challenging task. Our Skeleton Recall Loss \textit{demonstrates powerful multi-class segmentation capabilities} in addition to standard binary settings. \cref{tab:results_thin} showcases the results of multi-class segmentation on 13 different brain vessel classes of the TopCoW dataset using both standard nnUNet and our proposed Loss. The results demonstrate that while nnUNet exhibits slightly better $\beta_0$ error, our Skeleton Recall Loss significantly improves Dice and clDice scores. Moreover, it performs on par in terms of $\beta_1$ error, ultimately yielding a superior overall result.

\begin{figure}[t]
    \centering
    \includegraphics[width=0.49\textwidth]{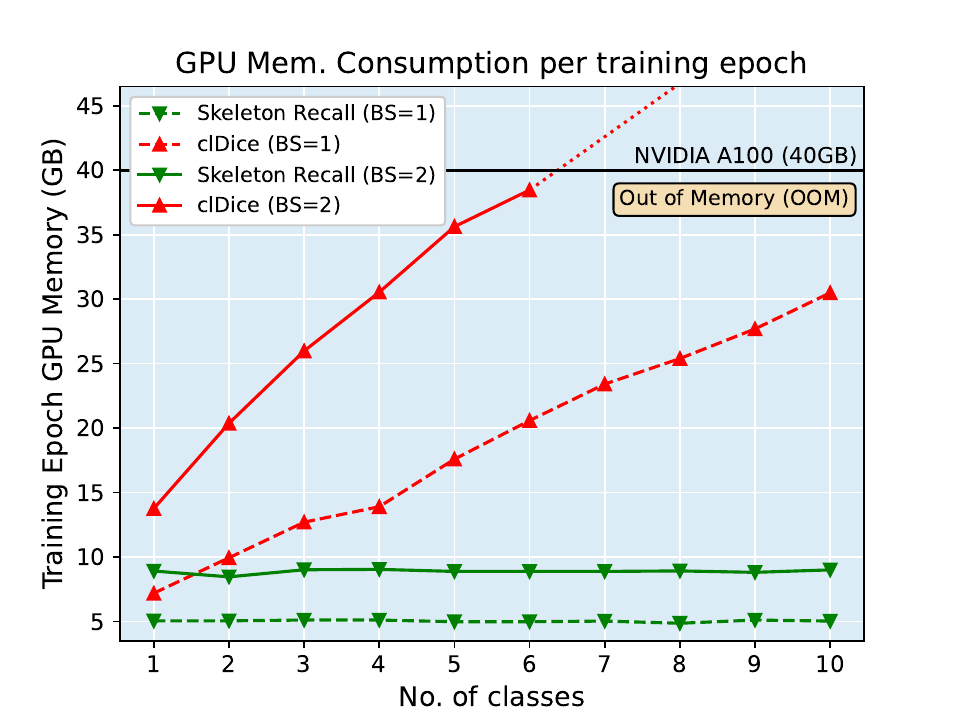}
    \includegraphics[width=0.49\textwidth]{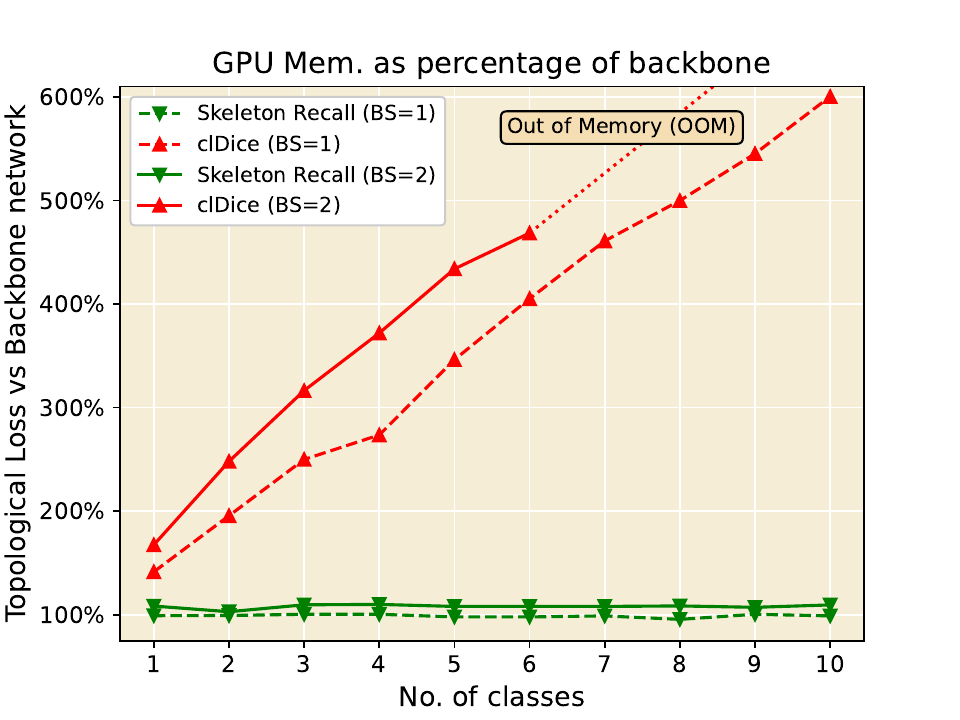}
    \includegraphics[width=0.49\textwidth]{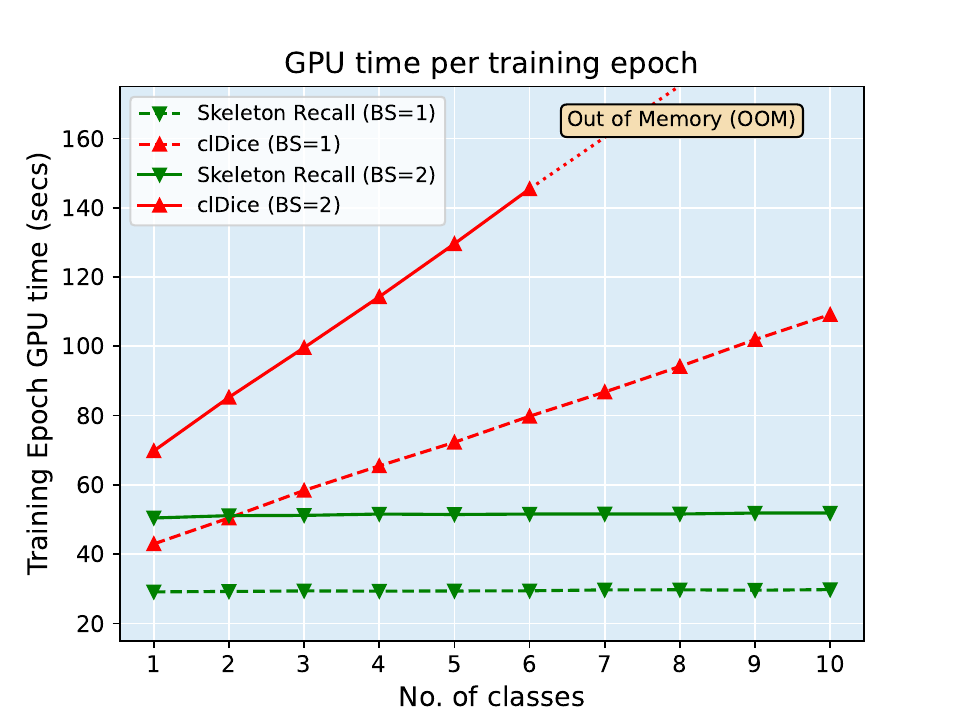}
    \includegraphics[width=0.49\textwidth]{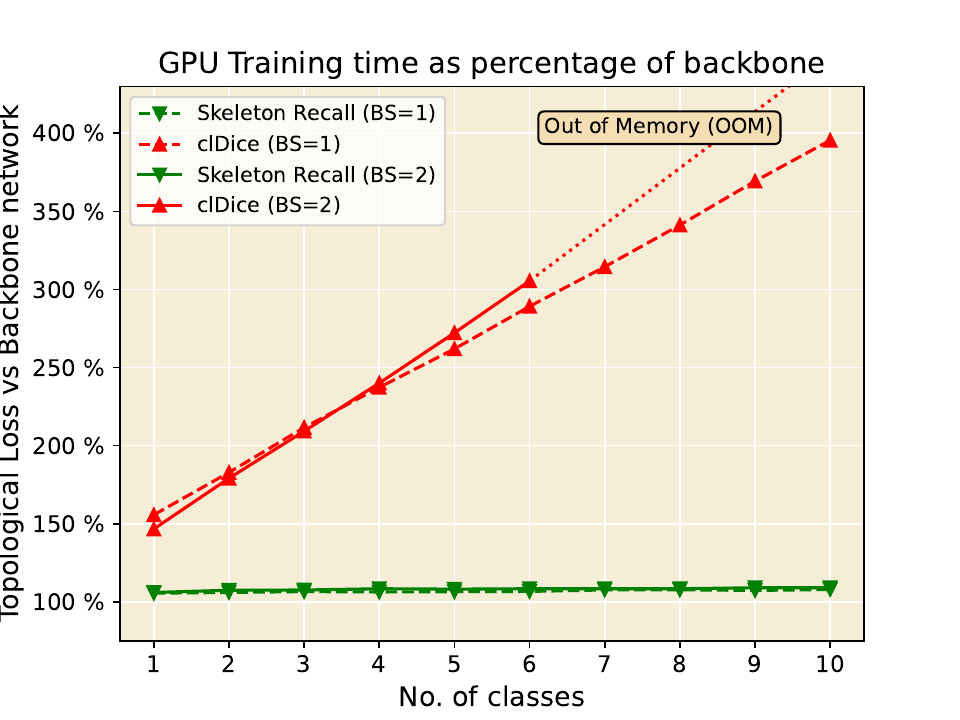}
    \caption{\textbf{Resource Utilization for Multi-class Segmentation.} \textit{Skeleton Recall Loss} requires minimal additional GPU memory and training time for an increasing number of classes, as it avoids a differentiable skeleton computation. Competing methods like \textit{clDice Loss}, on the other hand, incur enormous overheads which make multi-class training on datasets such as TopCoW \cite{topcow} infeasible. Analysis was performed for two different batch sizes (BS) on a single A100 40GB GPU, averaged over 5 training epochs, for ease of comparability.}
    \label{fig:mem_time_plots}
\end{figure}

\cref{fig:mem_time_plots} shows the multi-class resource utilization with respect to the number of classes of our proposed Loss in comparison to clDice Loss. We demonstrate significant training time and memory savings with near-constant additional overhead despite the increasing number of classes. In contrast, the plots underscore the approximately linear growth in memory consumption and training time associated with clDice Loss. We note that the inefficiency of clDice Loss rendered it infeasible on all 13 classes as it exceeded the memory capacity of an A100 40GB GPU. In summary, Skeleton Recall Loss can be employed for an arbitrary number of classes with minimal computational cost.

\section{Conclusion}

This paper proposes a novel loss function, \textit{Skeleton Recall Loss}, designed for connectivity preserving semantic segmentation. It is domain and architecture agnostic and, unlike existing methods, requires minimal additional training time and memory. Through extensive evaluation on five publicly available datasets, we demonstrate that Skeleton Recall Loss shows overall superior performance on existing state-of-the-art topology-aware loss functions. Moreover, it stands as the first loss function designed for computationally manageable thin structure segmentation within the increasingly significant but hitherto unaddressed multi-class context. In essence, Skeleton Recall Loss represents a significant advancement in the field of thin structure segmentation, offering both efficiency and efficacy. The public availability of our code further facilitates its adoption and serves as a foundation for future advancements in this critical area of study.

\section*{Acknowledgement}
The present contribution is supported by the Helmholtz Association under the joint research school "HIDSS4Health – Helmholtz Information and Data Science School for Health". This work was partly funded by Helmholtz Imaging (HI), a platform of the Helmholtz Incubator on Information and Data Science. PV is funded through an Else Kröner Clinician Scientist Endowed Professorship by the Else Kröner Fresenius Foundation (reference number: 2022\_EKCS.17).

\newpage

\bibliographystyle{splncs04}
\bibliography{references}

\newpage

\appendix

\section{Influence of the Loss Weight Parameter $\bm{w}$}

\begin{figure}[ht]
    \centering
    \includegraphics[width=0.49\textwidth]{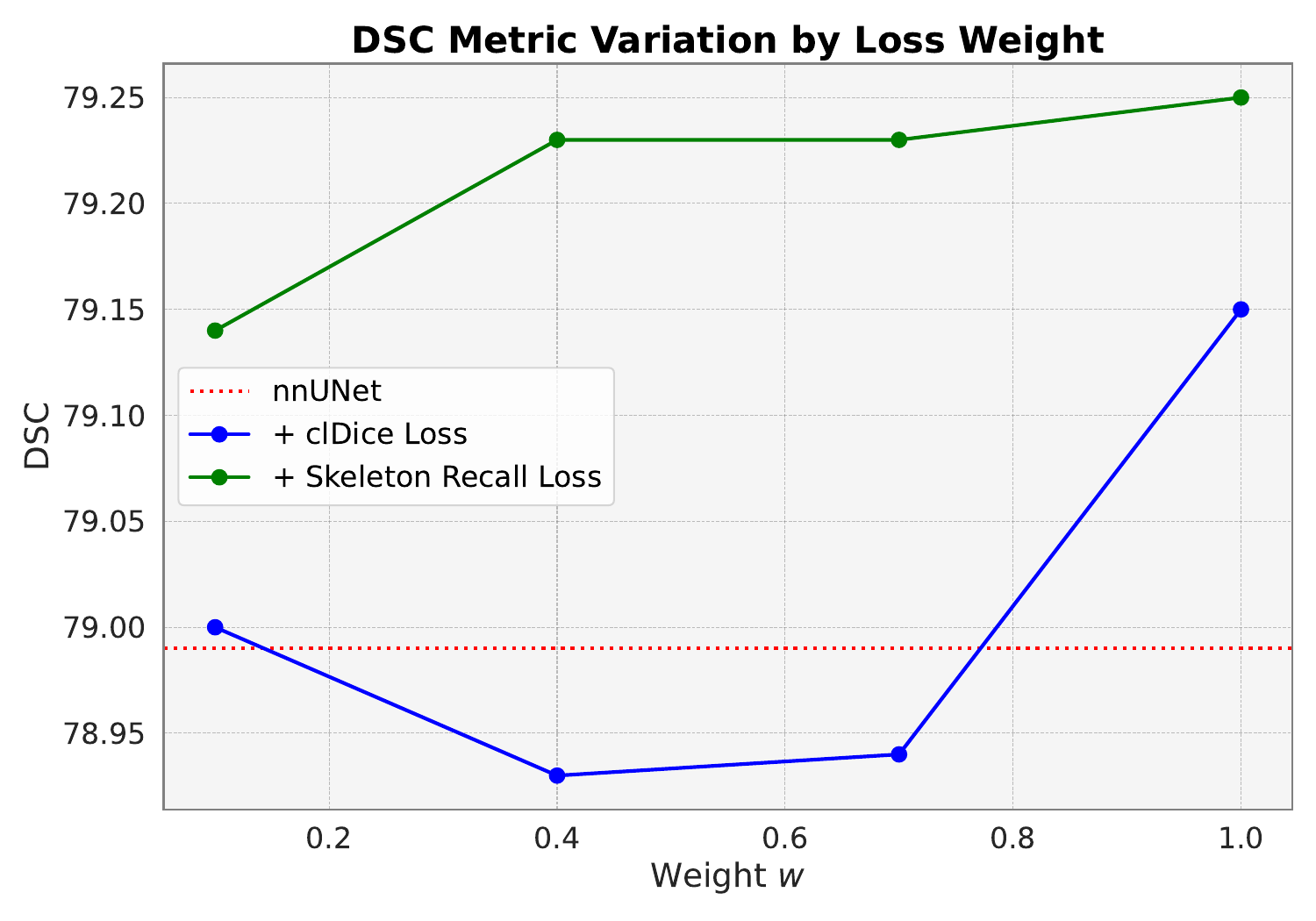}
    \includegraphics[width=0.49\textwidth]{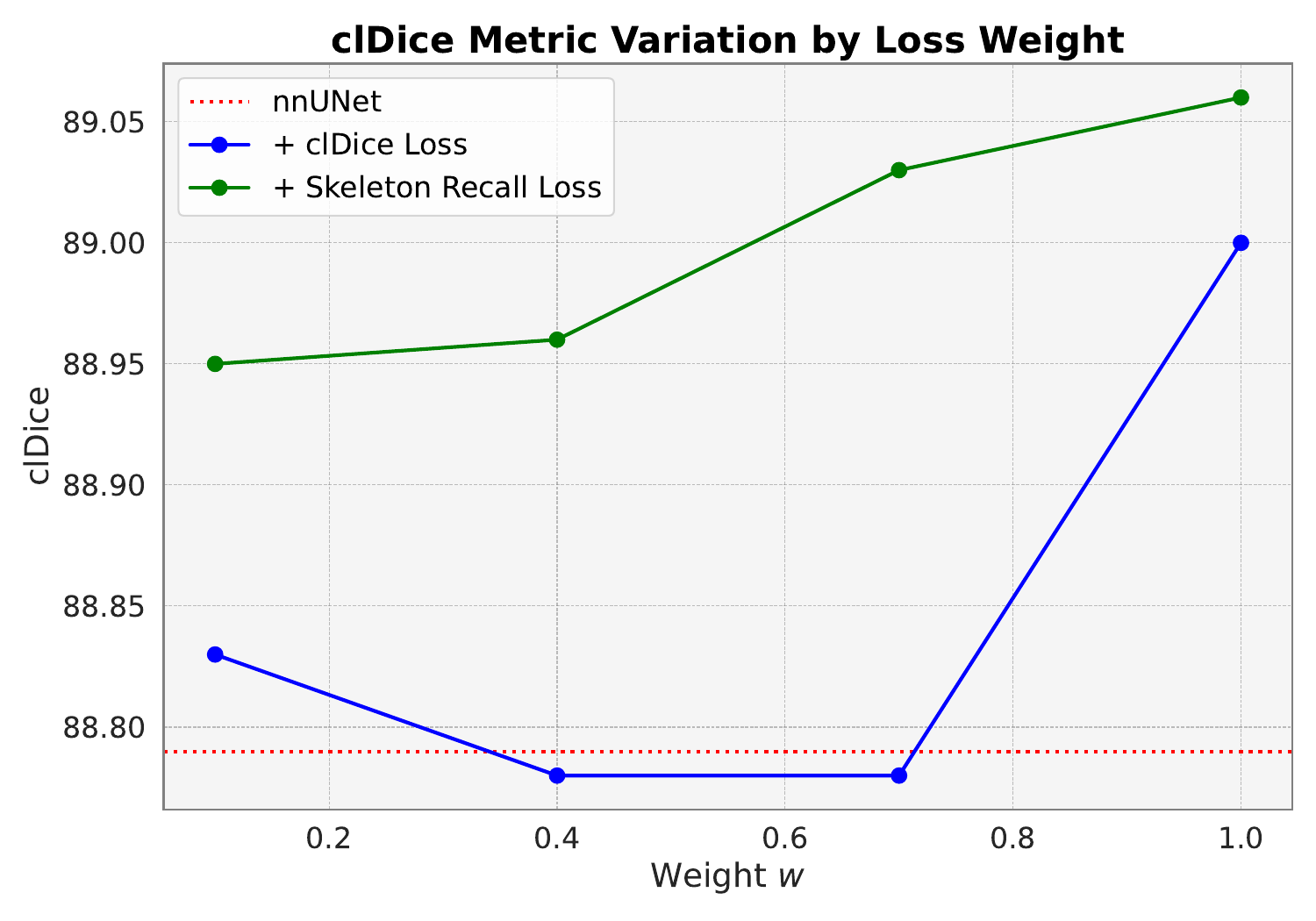}
    \caption{\textbf{Evaluation of weight parameter} $\bm{w}$: The nnUNet baseline performance on Roads is depicted in \textit{red}. Altering the weight $w$ influences the impact of the additional loss. The figure shows that our Skeleton Recall Loss (\textit{green}) consistently surpasses clDice Loss (\textit{blue}) irrespective of the weight parameter.}
    \label{fig:weight_variation}
\end{figure}

\section{Model Configurations}

\begin{table}[ht]
  \aboverulesep=0ex
  \belowrulesep=0ex
  \caption{\textbf{Configuration of nnUNet and HRNet on the five datasets:} nnUNet employs patch-based training and inference, while HRNet uses the whole image. HRNet is designed specifically for 2D data, while nnUNet supports both 2D and 3D images.}
  \label{tab:network_config}
  \centering
  \begin{adjustbox}{width=0.95\textwidth}
  \begin{tabular}{@{\hspace{1mm}}l@{\hspace{1mm}}l@{\hspace{1mm}}|@{\hspace{1mm}}c@{\hspace{1mm}}c@{\hspace{4mm}}c@{\hspace{1mm}}c@{\hspace{1mm}}}
    \toprule
    \rule{0pt}{1.1EM}
    \textbf{Dataset} & \textbf{Network} & \textbf{Batch Size} & \textbf{Patch Size} & \textbf{Optimizer} & \textbf{LR Schedule}\\
    \toprule
    \rule{0pt}{1.1EM}
    \multirow{2}{*}{\shortstack{Roads}} & nnUNet & 12 & $512\!\times\!512$ & SGD, $\mu=0.99$ & PolyLR(1e-2)\\
     & HRNet & 2 & $1500\!\times\!1500$ & SGD, $\mu=0.9$ & PolyLR(1e-2) \\ 
    \midrule
    \rule{0pt}{1.1EM}
    \multirow{2}{*}{\shortstack{DRIVE}} & nnUNet & 2 & $512\!\times\!512$ & SGD, $\mu=0.99$ & PolyLR(1e-2)\\
     & HRNet & 2 & $565\!\times\!584$ & SGD, $\mu=0.9$ & PolyLR(1e-2)\\ 
    \midrule
    \rule{0pt}{1.1EM}
    \multirow{2}{*}{\shortstack{Cracks}} & nnUNet & 65 & $224\!\times\!224$ & SGD, $\mu=0.99$ & PolyLR(1e-2)\\
     & HRNet & 64 & $224\!\times\!224$ & SGD, $\mu=0.9$ & PolyLR(1e-2)\\ 
    \midrule
    \rule{0pt}{1.1EM}
    \multirow{2}{*}{\shortstack{ToothFairy}} & nnUNet & 2 & $80\!\times\!160\!\times\!192$ & SGD, $\mu=0.99$ & PolyLR(1e-2)\\
     & HRNet & -- & -- & -- & --\\ 
    \midrule
    \rule{0pt}{1.1EM}
    \multirow{2}{*}{\shortstack{TopCoW}} & nnUNet & 2 & $80\!\times\!192\!\times\!160$ & SGD, $\mu=0.99$ & PolyLR(1e-2)\\
     & HRNet & -- & -- & -- & --\\ 
  \bottomrule
  \end{tabular}
  \end{adjustbox}
\end{table}

\end{document}